\documentclass[12pt]{article}

\input epsf
\usepackage{amssymb,amsmath,wrapfig,subfigure}
\usepackage[matrix,arrow,curve]{xy}
\usepackage{cite}
\usepackage{graphicx,color}
\usepackage[debug,pageanchor=false]{hyperref}
\definecolor{link}{rgb}{.8,.15,.1}
\hypersetup{colorlinks=true,linkcolor=link,citecolor=link,urlcolor=link,linktocpage}
\usepackage{cancel}
\usepackage{comment}

\makeatletter
\@addtoreset{equation}{section}
\makeatother

\def\rr {{\mathbb{R}}}

\def\del {\partial}

\def\del {\partial}

\def\vol {\mathrm{vol}}

\def\lsim{\mathrel{\rlap{\lower4pt\hbox{\hskip1pt$\sim$}}
    \raise1pt\hbox{$<$}}}                

\def\Re{\mathrm{Re}}
\def\Im{\mathrm{Im}}
\def\ads{\mathrm{AdS}_6 \times M_4}

\newcommand\cl[2]{\mathrm{Cl}({#1},{#2})} 
\newcommand\pai[2]{\left( {#1}, {#2} \right)} 
\newcommand\ep[2]{e_{{#1}_{#2}}} 

\begin{document}

\begin{titlepage}

\begin{center}

\vskip .3in \noindent

{\Large \bf{AdS$_6$ solutions of type II supergravity}}

\bigskip

	Fabio Apruzzi$^1$, Marco Fazzi$^2$, Achilleas Passias$^3$, Dario Rosa$^3$ and Alessandro Tomasiello$^3$\\

       \bigskip
	 {\small $^1$ Institut f\"ur Theoretische Physik, Leibniz Universit\"at Hannover,
	Appelstra\ss e 2, 30167 Hannover, Germany \\
	 \vspace{.1cm}
	 $^2$ Physique Th\'eorique et Math\'ematique, Universit\'e Libre de Bruxelles, Campus Plaine C.P.~231, B-1050 Bruxelles, Belgium \\ and \\ International Solvay Institutes, Bruxelles, Belgium \\
	\vspace{.1cm} 
	$^3$ Dipartimento di Fisica, Universit\`a di Milano--Bicocca, Piazza della Scienza 3, I-20126 Milano, Italy\\
    and\\
    INFN, sezione di Milano--Bicocca
	}

       \vskip .5in
       {\bf Abstract }
       \vskip .1in
\end{center}

Very few AdS$_6 \times M_4$ supersymmetric solutions are known: one in massive IIA, and two IIB solutions dual to it. The IIA solution is known to be unique; in this paper, we use the pure spinor approach to give a classification for IIB supergravity. We reduce the problem to two PDEs on a two-dimensional space $\Sigma$. 
$M_4$ is then a fibration of $S^2$ over $\Sigma$; the metric and fluxes are completely determined in terms of the solution to the PDEs. The results seem likely to accommodate near-horizon limits of $(p,q)$-fivebrane webs studied in the literature as a source of CFT$_5$'s. We also show that there are no AdS$_6$ solutions in eleven-dimensional supergravity.

\noindent

\vfill
\eject

\end{titlepage}

\hypersetup{pageanchor=true}

\tableofcontents

\section{Introduction} 
\label{sec:intro}

One of the interesting theoretical results of string theory is that it helps defining several nontrivial quantum field theories in dimensions higher than four, which are hard to study with traditional methods. For example, several five-dimensional superconformal field theories (SCFT$_5$'s) have been defined, using D4-branes in type I' \cite{seiberg-5d,morrison-seiberg}, M-theory on Calabi--Yau manifolds with shrinking cycles \cite{morrison-seiberg,intriligator-morrison-seiberg}, $(p,q)$-fivebrane webs \cite{aharony-hanany} (sometimes also including $(p,q)$-sevenbranes \cite{dewolfe-hanany-iqbal-katz}). These various realizations are dual to each other \cite{leung-vafa,dewolfe-hanany-iqbal-katz}; some of these theories are also related by compactification \cite{benini-benvenuti-tachikawa} to the four-dimensional ``class S'' theories \cite{gaiotto}.

However, not too many AdS$_6$ duals are known to these SCFT$_5$'s. Essentially the reason is that there is no D-brane stack whose near-horizon limit gives AdS$_6$. Indeed the string realizations quoted above originate from intersecting branes, whose localized metrics are notoriously difficult to find, as illustrated for example in \cite{lunin}; even were they known, the relevant near-horizon limit would probably be far from obvious. One exception is when one of the branes is completely inside the other; in such cases some partially delocalized solutions \cite{youm} become actually localized. This was used by Brandhuber and Oz \cite{brandhuber-oz} to obtain the first AdS$_6$ solution in string theory. (It was also anticipated to exist \cite{ferrara-kehagias-partouche-zaffaroni-ads6} as a lift of a vacuum in the six-dimensional supergravity of \cite{romans-F4}.) It is in massive IIA, and it represents the near-horizon limit of a stack of D4's near an O8--D8 wall; thus it is dual to the theories in \cite{seiberg-5d}. The internal space is half an $S^4$; the warping function $A$ and the dilaton $\phi$ go to infinity at its boundary. This is just a consequence of the presence of the O8--D8 system there, and it is a reflection of the peculiar physics of the corresponding SCFT$_5$'s. The fact that the dilaton diverges at the wall roughly corresponds to a Yang--Mills kinetic term of the type $\phi F_{\mu \nu} F^{\mu\nu}$; the scalar $\phi$ plays the role of $\frac1{g_{\rm YM}^2}$, and at the origin $\phi\to 0$ one finds a strongly coupled fixed point. 

One can also study a few variations on the Brandhuber--Oz solution, such as orbifolding it \cite{bergman-rodriguezgomez} and performing T-duality \cite{cvetic-lu-pope-vazquezporitz,lozano-colgain-rodriguezgomez-sfetsos} or even the more recently developed \cite{sfetsos-thompson,lozano-colgain-sfetsos-thompson} nonabelian T-duality \cite{lozano-colgain-rodriguezgomez-sfetsos,lozano-colgain-rodriguezgomez}. The latter is not thought to be an actual duality, but rather a solution-generating duality; thus the solution should represent some new physics, although its global features are puzzling \cite{lozano-colgain-rodriguezgomez}.

In this paper, we attack the problem systematically, using the ``pure spinor'' techniques, emboldened by the recent success of this method for AdS$_7$ solutions of type II supergravity \cite{afrt}. In general, the procedure reformulates the equations for preserved supersymmetry in terms of certain differential forms defining $G$-structures on the ``generalized tangent bundle'' $T\oplus T^*$. It originates from generalized complex geometry \cite{hitchin-gcy,gualtieri} and its first application was to Minkowski$_4$ or AdS$_4 \times M_6$ solutions of type II supergravity \cite{gmpt2}, in which case the relevant $G$ was ${\rm SU}(3) \times {\rm SU}(3)$. In \cite{10d} the method was extended (still in type II supergravity) to any ten-dimensional geometry; in this paper we apply to AdS$_6\times M_4$ the general system obtained there. We work in IIB, since in massive IIA the Brandhuber--Oz solution is unique \cite{passias}, and in eleven-dimensional supergravity there are no solutions, as we show in appendix \ref{app:11d}.

As in \cite{afrt}, the relevant structure on $T\oplus T^*$ is an ``identity'' structure (in other words, $G$ is the trivial group). Such a structure is defined by a choice of two vielbeine $e^a_\pm$ (roughly associated with left- and right-movers in string theory). Just as in \cite{afrt}, we actually prefer working with a single ``average'' vielbein $e^a$ and with some functions on $M_4$ encoding the map between the two vielbeine $e^a_\pm$. We then use these data to parameterize the forms appearing in the supersymmetry system. The supersymmetry equations then determine $e^a$ in terms of the functions on $M_4$, thus also determining completely the local form of the metric. As usual for this kind of formalism, the fluxes also come out as an output; less commonly, but again just as in \cite{afrt}, the Bianchi identities are automatically satisfied. 

When the dust settles, it turns out that we have completely reduced the problem to a system of two PDEs (see (\ref{eq:dz}), (\ref{eq:a2}) below) on a two-dimensional space $\Sigma$. The metric is that of an $S^2$-fibration over $\Sigma$. This should not come as a surprise: a SCFT$_5$ has an SU$(2)$ R-symmetry, which manifests itself in the gravity dual as the isometry group of the $S^2$. In \cite{afrt}, for similar reasons the internal space $M_3$ was an $S^2$-fibration over an interval. 

In AdS$_7$ the problem was reduced in \cite{afrt} to a system of first-order ODEs, which was then easy to study numerically; in our present case of supersymmetric AdS$_6$ solutions, we have PDEs, which are harder to study even numerically. Using EDS techniques (see for example \cite[Chap.~III]{bryant-chern-gardner-goldshmidt-griffiths} or \cite[Sec.~10.4.1]{stephani-kramer-maccallum-hoenselaers-herlt}) we have checked that the system is ``well-formed'': the general solution is expected to depend on two functions of one variable, which can be thought of as the values of the warping function $A$ and the dilaton $\phi$ at the boundary of $\Sigma$. (We expect regularity of the metric to fix those degrees of freedom as well, up to discrete choices.) We do recover two explicit solutions to the PDEs, corresponding to the abelian and nonabelian T-duals of the Brandhuber--Oz solution mentioned above.

Even though we do not present any new solutions in this paper, it seems likely that our PDEs will describe $(p,q)$-fivebrane webs. For the AdS$_7$ case, it was conjectured \cite{gaiotto-t-6d} that the new solutions found in \cite{afrt} arise as near-horizon limits of NS5--D6--D8 configurations previously studied in \cite{hanany-zaffaroni-6d,brunner-karch}. The fact that those solutions have cohomogeneity one (namely, that all fields only depend on the coordinate on the base interval) matches with the details of the configuration. The coordinates $x^0,\ldots, x^5$ are common to all branes; the NS5's are located at $x^7=x^8=x^9=0$, while their positions in $x^6$ parameterize the tensor branch of the SCFT$_6$; the D6's are located at $x^7=x^8=x^9=0$, and extended along $x^6$; the D8's are extended along $x^7$, $x^8$, $x^9$, and located at various $x^6=x^6_{{\rm D8}_i}$. 

For AdS$_6$, the natural analogue of this story would involve $(p,q)$-fivebranes whose common directions would be $x^0,\ldots, x^4$, and which would be stretched along a line in the $x^5$--$x^6$ plane (such that $\frac{x^5}{x^6}=\frac pq$). It is natural to conjecture that the solutions to our PDEs would correspond to near-horizon limits of such configurations, with the $x^5$--$x^6$ plane somehow corresponding to our $\Sigma$; the remaining directions $x^7$, $x^8$, $x^9$ would provide our $S^2$ (as well as the radial direction of AdS$_6$). For such cases we would expect $\Sigma$ to have a boundary, at which the $S^2$ shrinks; the $(p,q)$-fivebranes would then be pointlike sources at this boundary.
We hope to come back on this in the near future.

The paper is organized as follows. In section \ref{sec:psp} we present the system (\ref{eq:64}) of differential equations for supersymmetry, expressed in terms of differential forms $\Phi$ and $\Psi$ describing an identity structure on $M_4$; the derivation from \cite{10d} is given in appendix \ref{app:10-6}. In section \ref{sec:para} we parameterize the differential forms in terms of a vielbein on $M_4$ and of four functions. We then plug this parameterization in the system, and obtain in section \ref{sec:gen} our results on the metric and fluxes, and the two PDEs (\ref{eq:ddz}), (\ref{eq:a2}) that one needs to satisfy. Finally, in section \ref{sec:pde}, we make some general remarks about the PDEs, and recover the known examples.


\section{Supersymmetry and pure spinor equations for AdS$_6$} 
\label{sec:psp}

We will start by presenting the system of pure spinor equations that we need to solve. Although this is similar to systems in other dimensions, there are some crucial differences, which we will try to highlight. 

The original example of the pure spinor approach to supersymmetry was found for Mink$_4 \times M_6$ or AdS$_4 \times M_6$ solutions in type II supergravity \cite{gmpt2}, where the BPS conditions were reformulated in terms of certain differential equations on an ${\rm SU}(3) \times {\rm SU}(3)$ structure on the ``generalized tangent bundle'' $TM_6\oplus T^*M_6$. Other examples followed over the years; for instance, \cite{lust-patalong-tsimpis} applied the strategy to Mink$_d \times M_{10-d}$ for even $d$ (for $d=2$ the situation was improved in \cite{prins-tsimpis,rosa,prins-tsimpis2}); the case $\rr \times M_9$ was considered in \cite{koerber-martucci-ads}. 

Partially motivated by the need of generating quickly pure-spinor-like equations for different setups, \cite{10d} formulated a system directly in ten dimensions, using the geometry of the generalized tangent bundle of $M_{10}$. This could have also been used in \cite{afrt} to generate a system for AdS$_7\times M_3$ solutions in type II; in that case, however, it was more convenient to derive the system from the one of \cite{lust-patalong-tsimpis} for Mink$_6 \times M_4$, via a cone construction. This approach is not as readily available for our current case AdS$_6 \times M_4$; hence, we will attack it directly from \cite{10d}. 

We describe the derivation of our system from the ten-dimensional one of \cite{10d} in appendix \ref{app:10-6}. The system in \cite{10d} contains two ``symmetry'' equations (3.1b) that usually simply fix the normalizations of the pure spinors; two ``pairing'' equations (3.1c,d) that often end up being redundant (although not always, see \cite{rt,rosa}); and one  ``exterior'' equation (3.1a) that usually generates the pure spinor equations one is most interested in. This pattern is repeated for our case. One important difference is that the spinor decomposition we have to start with is clumsier than the one in other dimensions. Usually, the ten-dimensional spinors $\epsilon_a$ are the sum of two (or sometimes even one) tensor products. For AdS$_4\times M_6$ in IIB, for example, we simply have $\epsilon_a = \zeta_{4\,+} \otimes \eta^a_{6\,+} + {\rm c.c.}$.  The analogue of this for Mink$_6\times M_4$ in IIB would be 
\begin{equation}\label{eq:epsmink}
	\begin{split}
	\epsilon_1 = \zeta_{6\,+} \otimes \eta^1_{4\,+} + \zeta_{6\,+}^c \otimes \eta^{1\,c}_{4\,+}\\
	\epsilon_2 = \zeta_{6\,+} \otimes \eta^2_{4\,\mp} + \zeta_{6\,+}^c \otimes \eta^{2\,c}_{4\,\pm}\\
	\end{split}
		 \qquad ({\rm Mink}_6 \times M_4;\ {\rm IIA/IIB})\ ,
\end{equation}
where $(\ )^c \equiv C (\ )^* $ denotes Majorana conjugation. For AdS$_6\times M_4$, however, such an Ansatz cannot work: compatibility with the negative cosmological constant of AdS$_6$ demands that the $\zeta_6$ obey the Killing spinor equation on AdS$_6$,
\begin{equation}\label{eq:KSE}
	\nabla_\mu \zeta_6 = \frac{1}{2}\gamma^{(6)}_\mu \zeta_6\ ,
\end{equation}
and solutions to this equation cannot be chiral, while the $\zeta_{6\,+}$ in (\ref{eq:epsmink}) are chiral. This issue does not arise in AdS$_4$ because in that case $(\zeta_{4\,+})^c$ has negative chirality; here $(\zeta_{6\,+})^c$ has positive chirality. This forces us to add ``by hand'' to (\ref{eq:epsmink}) a second set of spinors with negative chirality, ending up with the unpromising-looking
\begin{equation}\label{eq:10deps}
	\begin{split}
	\epsilon_1 = \zeta_+ \eta^1_+ + \zeta_+^c {\eta^1_+}^c + \zeta_- \eta^1_- + \zeta_-^c {\eta^1_-}^c \\
	\epsilon_2 = \zeta_+ \eta^2_\mp + \zeta_+^c {\eta^2_\mp}^c + \zeta_- \eta^2_\pm + \zeta_-^c {\eta^2_\pm}^c\\
	\end{split}
		 \qquad ({\rm AdS}_6 \times M_4;\ {\rm IIA/IIB})\ 
\end{equation}
where we have dropped the ${}_6$ and ${}_4$ labels (and the $\otimes$ sign), as we will do elsewhere. Attractive or not, (\ref{eq:10deps}) will turn out to be the correct one for our classification.

In the main text from now on we will consider the IIB case (unless otherwise stated). This is because AdS$_6 \times M_4$ solutions in massive IIA were already analyzed in \cite{passias}, where it was found that the only solution is the one in \cite{brandhuber-oz}. We did find it useful to check our methods on that solution as well; we sketch how that works in appendix \ref{app:IIA}. As for the massless case, we found it more easily attacked by direct analysis in eleven-dimensional supergravity, which we present in appendix \ref{app:11d}, given that it is methodologically a bit outside the stream of our pure spinor analysis in IIB.

With the spinor Ansatz (\ref{eq:10deps}) in hand, we can apply the system in \cite{10d}; the details of the derivation are described in appendix \ref{app:10-6}. We first describe the forms appearing in the system. If we were interested in the Minkowski case, the system would only contain the bispinors $\eta^1_+ \otimes \eta_+^{2\,\dagger}$ and $\eta^1_+ \otimes (\eta_+^{2\,c})^\dagger$.\footnote{\label{foot:cliff}As usual, we will identify forms with bispinors via the Clifford map $dx^{m_1}\wedge \ldots \wedge dx^{m_k}\mapsto \gamma^{m_1\ldots m_k}$.} (As usual in the pure spinor approach, we need not consider spinors of the type e.g.~$\eta^1_+ \otimes \eta^{1\,\dagger}_+$ to formulate a system which is necessary and sufficient.) Mathematically, this would describe an ${\rm SU}(2) \times {\rm SU}(2)$ structure on $TM_4 \oplus T^*M_4$. Since in (\ref{eq:10deps}) we also have the negative chirality spinors $\eta^1_-$ and $\eta_-^{1\,c}$, there are many more forms we can build. We have the even forms:\footnote{Notice that the ${}^1$ or ${}^2$ on $\phi$ has nothing to do with the ${}^1$ or ${}^2$ on the $\eta$'s; rather, it has to do with whether the second spinor is Majorana conjugated (${}^2$) or not (${}^1$). Another caveat is that the ${}_\pm$ does not indicate the degree of the form, as it is often the case in similar contexts; all the $\phi$'s in (\ref{eq:phi}) are even forms. One can think of the ${}_\pm$ as indicating whether these forms are self-dual or anti-self-dual.}
\begin{subequations}\label{eq:phipsi}
	\begin{equation}\label{eq:phi}
		\phi^1_\pm = e^{-A}\eta^1_\pm \otimes \eta_\pm^{2\,\dagger} \ ,\qquad
		\phi^2_\pm = e^{-A}\eta^1_\pm \otimes (\eta_\pm^{2\,c})^\dagger \equiv e^{-A} \eta^1_\pm \otimes \overline{\eta^2_\pm}\ ;
	\end{equation}
	and the odd forms:
	\begin{equation}\label{eq:psi}
		\psi^1_\pm = e^{-A}\eta^1_\pm \otimes \eta_\mp^{2\,\dagger} \ ,\qquad
		\psi^2_\pm = e^{-A}\eta^1_\pm \otimes (\eta_\mp^{2\,c})^\dagger \equiv e^{-A} \eta^1_\pm \otimes\overline{\eta^2_\mp}\ .
	\end{equation}
\end{subequations}
The factors $e^{-A}$ are inserted so that the bispinors have unit norm, in a sense to be clarified shortly; $A$ is the warping function, defined as usual by 
\begin{equation}\label{eq:A}
	ds^2_{10} = e^{2A} ds^2_{{\rm AdS}_6} + ds^2_{M_4}\ .
\end{equation}  
Already by looking at (\ref{eq:phi}), we see that we have \emph{two} ${\rm SU}(2) \times {\rm SU}(2)$ structures on $TM_4 \oplus T^*M_4$. If both of these structures come for example from ${\rm SU}(2)$ structures on $TM_4$, we see that we get an identity structure on $TM_4$, i.e.~a vielbein. In fact, this is true in general: (\ref{eq:phi}) always defines a vielbein on $M_4$. We will see in section \ref{sec:para} how to parameterize both (\ref{eq:phi}) and (\ref{eq:psi}) in terms of the vielbein they define. 

In the meantime, we can already now notice that the (\ref{eq:phi}) and (\ref{eq:psi}) can be assembled more conveniently using the SU(2) R-symmetry. This is the group that rotates ${\zeta\choose \zeta^c}$ and each of ${\eta^a_\pm \choose \eta^{a\,c}_\pm}$ as a doublet. One can check that (\ref{eq:10deps}) is then left invariant, so it is a symmetry; since it acts on the external spinors, we call it an R-symmetry. It is the manifestation of the R-symmetry of a five-dimensional SCFT. Something very similar was noticed in \cite{afrt} for AdS$_7$: the pure spinor system ((2.11) in that paper) naturally assembled into singlets and one triplet of SU(2). (Recall that a six-dimensional SCFT also has an SU(2) R-symmetry.) While in that paper the SU(2) formalism was only stressed at the end of the computations, here the analysis is considerably more complicated, and SU(2) will be used from the very beginning to yield more manageable results. Let us define
\begin{subequations}\label{eq:su2}
\begin{align}
\Phi_\pm &\equiv 
	\begin{pmatrix} \eta^1_\pm \\ \eta^{1\,c}_\pm \end{pmatrix} 
		\otimes 
	\begin{pmatrix} \eta^{2\,\dagger}_\pm & \overline{\eta^2_\pm} \end{pmatrix} = 				
	\begin{pmatrix} \phi^1_\pm & \phi^2_\pm \\ -(\phi^2_\pm)^* & (\phi^1_\pm)^*\end{pmatrix} \nonumber\\
	&= \Re \phi^1_\pm \mathrm{Id}_2 +i (\Im \phi^2_\pm \sigma_1 +\Re \phi^2_\pm\sigma_2 +\Im \phi^1_\pm \sigma_3) 
	\equiv \Phi_{\pm}^0\mathrm{Id}_2 +i \Phi_{\pm}^\alpha \sigma_\alpha\ , \\
\Psi_\pm &\equiv 
	\begin{pmatrix} \eta^1_\pm \\ \eta^{1\,c}_\pm \end{pmatrix} 
		\otimes 
	\begin{pmatrix} \eta^{2\,\dagger}_\mp & \overline{\eta^2_\mp} \end{pmatrix} = 		
	\begin{pmatrix} \psi^1_\pm & \psi^2_\pm \\ -(\psi^2_\pm)^* & (\psi^1_\pm)^* \end{pmatrix}\nonumber\\
	 &= \Re \psi^1_\pm \mathrm{Id}_2+i(\Im \psi^2_\pm \sigma_1 +\Re \psi^2_\pm\sigma_2 +\Im \psi^1_\pm \sigma_3) 
	\equiv \Psi^0_\pm \mathrm{Id}_2 + i \Psi_{\pm}^\alpha \sigma_\alpha\ .
\end{align}
\end{subequations}
$\sigma_\alpha$, $\alpha=1,2,3$, are the Pauli matrices. Here and in what follows, the superscript ${}^0$ denotes an SU(2) singlet, and not the zero-form part; the superscript ${}^\alpha$ denotes an SU(2) triplet, not a one-form. We hope this will not create confusion.

As we already mentioned, the forms $\Phi_\pm$, $\Psi_\pm$ will define an identity structure on $M_4$. However, not any random forms $\Phi_\pm$, $\Psi_\pm$ may be written as bispinors as in (\ref{eq:su2}). In other cases, such as for ${\rm SU}(3) \times {\rm SU}(3)$ structures in six dimensions \cite{gmpt2}, it is useful to formulate a set of constraints on the forms that guarantee that they come from spinors; this allows to completely forget about the original spinors, and formulate supersymmetry completely in terms of some forms satisfying some constraints. In the present case, it would be possible to set up such a fancy approach, by saying that $\Phi_\pm$ and $\Psi_\pm$ should satisfy a condition on their inner products. For example we could impose that the $\Phi$'s and $\Psi$'s be pure spinors on $M_4$ obeying the compatibility conditions\footnote{The Chevalley--Mukai pairing is defined as $(\alpha, \beta)=(\alpha \wedge \lambda(\beta))_4$, where on a $k$-form $\lambda \omega_k = (-)^{\lfloor \frac k2 \rfloor} \omega_k$.\label{foot:lambda}}
\begin{equation}\label{eq:comp}
	(\Phi_{\pm}^\alpha,\Phi_{\pm}^\beta)= (\Psi_{\pm}^\alpha,\Psi_{\pm}^\beta)=
	\delta^{\alpha \beta}(\Phi_{\pm}^0,\Phi_{\pm}^0)=
	\delta^{\alpha \beta}(\Psi_{\pm}^0, \Psi_{\pm}^0)\ .
\end{equation}
As in \cite{afrt}, this would however be an overkill, since in section \ref{sec:para} we will directly parameterize $\Phi_\pm$ and $\Psi_\pm$ in terms of a vielbein and some functions on $M_4$. This will achieve the end of forgetting about the spinors $\eta^a_\pm$ by different means.

We can finally give the system of equations equivalent to preserved supersymmetry:
\begin{subequations}\label{eq:64}
\begin{align}
&d_H \left[e^{3A-\phi} (\Psi_- -\Psi_+)^0 \right] - 2e^{2A-\phi} (\Phi_- + \Phi_+)^0 = 0 \ , \label{eq:a}\\
&d_H \left[e^{4A-\phi} (\Phi_- -\Phi_+)^\alpha \right] - 3e^{3A-\phi} (\Psi_- + \Psi_+)^\alpha = 0 \ , \label{eq:bcd}\\
&d_H \left[e^{5A-\phi} (\Psi_- -\Psi_+)^\alpha \right] - 4e^{4A-\phi} (\Phi_- + \Phi_+)^\alpha = 0 \ , \label{eq:efg}\\
&d_H \left[e^{6A-\phi} (\Phi_- -\Phi_+)^0 \right] - 5e^{5A-\phi} (\Psi_- + \Psi_+)^0 = - \frac14 e^{6A} \ast_4 \lambda F \ , \label{eq:h}\\
&d_H \left[e^{5A-\phi} (\Psi_- + \Psi_+)^0 \right] = 0 \ ; \label{eq:i}\\
& || \eta^1 ||^2 = || \eta^2 ||^2 = e^A \ . \label{eq:norms}
\end{align}
\end{subequations}
As usual, $\phi$ here is the dilaton; $d_H = d - H\wedge$; $A$ was defined in (\ref{eq:A}); $\lambda$ is a sign operator defined in footnote \ref{foot:lambda}; $F= F_1+F_3$ is the ``total'' allowed internal RR flux, which also determines the external flux via
\begin{equation}\label{eq:10dflux}
	F_{(10)} = F+ e^{6A} {\rm vol}_6 \wedge *_4 \lambda F \ .
\end{equation} 
Again, we remind the reader that the superscript ${}^0$ denotes a singlet part, and ${}^\alpha$ a triplet part, as in (\ref{eq:su2}).

The last equation, (\ref{eq:norms}), can be reformulated in terms of $\Phi$ and $\Psi$. Since $\Vert \eta^a \Vert^2 \equiv \Vert \eta^a_+ \Vert^2 + \Vert \eta^a_- \Vert^2$, we can define $|| \eta^1_+ || = e^{A/2} \cos(\alpha/2)$, $|| \eta^1_- ||= e^{A/2} \sin(\alpha/2)$, $|| \eta^2_+ || = e^{A/2} \cos(\tilde \alpha/2)$, $|| \eta^2_- ||= e^{A/2} \sin(\tilde \alpha/2)$, where $\alpha,\tilde \alpha \in [0,\pi]$; we then get
\begin{equation}\label{eq:normsP}
\begin{split}
	(\Phi_+^0,\Phi_+^0)= \frac{1}{8} \cos^2(\alpha/2) \cos^2(\tilde \alpha/2) \ ,\qquad
	(\Phi_-^0,\Phi_-^0)= -\frac{1}{8} \sin^2(\alpha/2) \sin^2(\tilde \alpha/2) \ ; \\
	(\Psi_+^0,\Psi_-^0)= \frac{1}{8} \cos^2(\alpha/2) \sin^2(\tilde \alpha/2)	\ ,\qquad 	(\Psi_-^0,\Psi_+^0)= -\frac{1}{8} \sin^2(\alpha/2) \cos^2(\tilde \alpha/2)\ .
\end{split}
\end{equation}
Just as (\ref{eq:comp}), however, such a fancy formulation will be ultimately made redundant by our parameterization of $\Phi$ and $\Psi$ in section \ref{sec:para}, which will satisfy  (\ref{eq:comp}) automatically, and where we will take care to implement (\ref{eq:norms}), so that (\ref{eq:normsP}) will be satisfied too.

We can check immediately that (\ref{eq:64}) imply the equations of motion for the flux, by acting on (\ref{eq:h}) with $d_H$ and using (\ref{eq:i}). The equations of motion for the metric and dilaton are then satisfied (as shown in general in \cite{lust-tsimpis} for IIA, and in \cite{gauntlett-martelli-sparks-waldram-ads5-IIB} for IIB); the equations of motion for $H$ are also implied, since they are \cite{koerber-tsimpis} for Minkowski$_4$ compactifications (which include Minkowski$_5$ as a particular case, and hence also AdS$_6$ by a conical construction). We will see later that the Bianchi identities for $F$ and $H$ are also automatically satisfied for this case, as was the case for \cite{afrt}. 

It is also interesting to compare the system (\ref{eq:64}) with the above-mentioned system for Minkowski$_6$ in \cite{lust-patalong-tsimpis}. First of all the second summands in the left-hand side of (\ref{eq:a})--(\ref{eq:h}) implicitly come with a factor proportional to $\sqrt{-\Lambda}$ that we have set to one (since it can be reabsorbed in the warping factor $A$). To take the Mink$_6$ limit, we can imagine to restore those factors, and then take $\Lambda\to 0$. Hence all the second summands in the left-hand side of (\ref{eq:a})--(\ref{eq:h}) will be set to zero. This is not completely correct, actually, because implicit in (\ref{eq:a})--(\ref{eq:efg}) there are more equations, that one can get by acting on them with $d_H$ (before taking the $\Lambda \to 0$ limit); we have to keep these equations as well. So far the limit works in the same way as for taking the $\Lambda\to 0$ limit from AdS$_4$ to Minkowski$_4$ in \cite{gmpt2}. In the present case, however, there is one more thing to take into account. As we have seen, in the Minkowski$_6$ case the spinor Ansatz can be taken to be (\ref{eq:epsmink}) rather than the more complicated (\ref{eq:10deps}) we had to use for AdS$_6$. To go from (\ref{eq:10deps}) to (\ref{eq:epsmink}), we can simply set $\eta^1_-=0$ and $\eta^2_\pm=0$. This sets to zero some of our bispinors; for the IIB case on which we are focusing, it sets to zero everything but $\Phi_+$. This makes some of the equations disappear; some others become redundant. All in all, we are left with
\begin{equation}
	d_H (e^{2A-\phi}\Phi^0_+)= 0 \ ,\qquad
	d_H (e^{4A-\phi}\Phi^\alpha_+) = 0 \ ,\qquad
	d_H (e^{6A-\phi}\Phi^0_+)= -\frac14 e^{6A} *_4 \lambda F\ ,
\end{equation}
which is \cite[Eq.~(4.11)]{lust-patalong-tsimpis} in our SU(2)-covariant language. (In \cite{afrt}, this system was quoted in a slightly different way: the last equation was mixed with the first, to yield $e^\phi F=16 *_4 \lambda(dA\wedge \Phi^0_+)$.)

In summary, in this section we have presented the system (\ref{eq:64}), which is equivalent to preserved supersymmetry for backgrounds of the form AdS$_6\times M_4$. The forms $\Phi$ and $\Psi$ are not arbitrary: they obey certain algebraic constraints expressing their origin as spinor bilinears in (\ref{eq:su2}), (\ref{eq:phipsi}). We will now give the general solution to those constraints, and then proceed in section \ref{sec:gen} to analyze the system.


\section{Parameterization of the pure spinors} 
\label{sec:para}

We have introduced in section \ref{sec:psp} the even forms $\Phi_\pm$ and the odd forms $\Psi_\pm$ (see (\ref{eq:su2}), (\ref{eq:phi}), (\ref{eq:psi})). These are the main characters in the system (\ref{eq:64}), which is equivalent to preserved supersymmetry. Before we start using the system, however, we need to characterize what sorts of forms $\Phi_\pm$ and $\Psi_\pm$ can be: this is what we will do in this section. 

\subsection{Even forms} 
\label{sub:even}

We will first deal with $\Phi_\pm$. We will actually first focus on $\Phi_+$, and then quote the results for $\Phi_-$. The computations in this subsection are actually pretty standard, and we will be brief. 

Let us start with the case $\eta^1_+ = \eta^2_+ \equiv \eta_+$. Assume also for simplicity that $|| \eta_+ ||^2 = 1$. In this case the bilinears define an SU(2) structure:
\begin{equation}
	\eta_+ \eta_+^\dagger= \frac14 e^{-ij_+} \ ,\qquad \eta_+ \overline{\eta_+} = \frac14 \omega_+\ ,
\end{equation}
where the two-forms $j_+$, $\omega_+$ satisfy
\begin{equation}\label{eq:jo+}
	j_+ \wedge \omega_+ = 0 \ ,\qquad \omega_+^2 = 0 \ ,\qquad \omega_+ \wedge \overline{\omega_+} = 2 j_+^2 = - \vol_4\ .
\end{equation}
We can also compute 
\begin{equation}
	\eta^c_+ \eta^{c\,\dagger}_+ = \frac14 e^{i j_+} \ ,\qquad
	\eta^c_+ \eta^\dagger_+ = -\frac14 \overline{\omega_+}\ .
\end{equation} 

Let us now consider the case with two different spinors, $\eta^1_+ \neq \eta^2_+$; let us again assume that they have unit norm. We can define (in a similar way as in \cite{halmagyi-t})
\begin{equation}\label{eq:eet}
	\eta_{0+} = \frac12 (\eta^1_+ - i \eta^2_+) \ ,\qquad \tilde \eta_{0+} = \frac12 (\eta^1_+ + i \eta^2_+)\ .
\end{equation}
Consider now $a_+ = \eta^{2\,\dagger}_+ \eta^1_+ $, $b_+= \overline{\eta^2_+} \eta^1_+$. $\{\eta^2_+, \eta^{2\,c}_+\}$ is a basis for spinors on $M_4$; $a_+$, $b_+$ are then the coefficients of $\eta^1_+$ along this basis. Since $\eta^a_+$ have both unit norm, we have $|a_+|^2 + |b_+|^2=1$. By multiplying $\eta^a_+$ by phases, we can assume that $a_+$ and $b_+$ are for example purely imaginary, and we can then parameterize them as $a_+=-i \cos(\theta_+)$, $b_+ = i \sin (\theta_+)$. Going back to (\ref{eq:eet}), we can now compute their inner products: 
\begin{equation}
	\eta_{0+}^\dagger \eta_{0+}= \cos^2\left(\frac{\theta_+}2\right) \ ,\qquad
	\eta_{0+}^\dagger \tilde \eta_{0+} = 0 \ ,\qquad \overline{\eta_{0+}} \tilde \eta_{0+}  = \frac12 \sin(\theta_+) \ .
\end{equation}
From this we can in particular read off the coefficients of the expansion of $\tilde \eta_{0+}$ along the basis $\{ \eta_{0+}, \eta^c_{0+}\}$. This gives $\tilde \eta_{0+}=\frac1{|| \eta_{0+} ||^2}(\eta_{0+}^\dagger \tilde \eta_{0+} \eta_{0+} +
\overline{\eta_{0+}} \tilde \eta_{0+} \eta^c_{0+}) = \tan\left(\frac{\theta_+}2\right) \eta^c_{0+}$. Recalling (\ref{eq:eet}), and defining now $\eta_{0+} = \cos\left(\frac{\theta_+}2\right)\eta_{+}$, we get
\begin{equation}\label{eq:eta12}
	 \eta^1_+=  \cos\left(\frac{\theta_+}2\right) \eta_+ + \sin\left(\frac{\theta_+}2\right) \eta^c_+ \ ,\qquad
	 \eta^2_+=  i\left(\cos\left(\frac{\theta_+}2\right) \eta_+ - \sin\left(\frac{\theta_+}2\right) \eta^c_+ \right)\ .
\end{equation}
From this it is now easy to compute $\eta^1_+ \eta^{2\,\dagger}_+$ and $\eta^1_+ \overline{\eta^2_+}$. Recall, however, that in the course of our computation we have first fixed the norms and then the phases of $\eta^a_+$. The norms of the spinors we need in this paper are not one; they were actually already parameterized before (\ref{eq:normsP}), so as to satisfy (\ref{eq:norms}). The factor $e^A$, however, simplifies with the $e^{-A}$ in the definition (\ref{eq:phi}). Let us also restore the phases we earlier fixed, by rescaling $ \eta^1_\pm \to e^{i u_\pm} \eta^1_\pm$, $\eta^2_\pm \to e^{i t_\pm} \eta^2_\pm$. All in all we get
\begin{subequations}\label{eq:eta++}
\begin{align}
	\phi^1_+ &= \frac 14 \cos(\alpha/2) \cos(\tilde \alpha/2)e^{i(u_+-t_+)}\cos(\theta_+) \exp\left[-\frac1{\cos(\theta_+)}(i j_+ + \sin(\theta_+) {\rm Re} \omega_+)\right] \ ,\\
	\phi^2_+ &= \frac14 \cos(\alpha/2) \cos(\tilde \alpha/2)e^{i(u_++t_+)} \sin(\theta_+) \exp\left[\frac1{\sin(\theta_+)}(\cos(\theta_+) {\rm Re} \omega_+ + i {\rm Im}  \omega_+)\right]\ .
\end{align}
\end{subequations}
The formulas for $\phi^{1,2}_-$ can be simply obtained by changing $\cos(\alpha/2)\to \sin(\alpha/2)$, $\cos(\tilde \alpha/2)$ $\to \sin(\tilde \alpha/2)$, and ${}_+ \to {}_-$ everywhere. The only difference to keep in mind is that the last equation in (\ref{eq:jo+}) is now replaced with $\omega_- \wedge \overline{\omega_-} = 2 j_-^2 = \vol_4$. 


\subsection{Odd forms} 
\label{sub:odd}
We now turn to the bilinears of ``mixed type'', i.e.~the $\psi^{1,2}_\pm$ we defined in (\ref{eq:psi}), which result in odd forms. We will again start from the case where $\eta^1_\pm = \eta^2_\pm \equiv \eta_\pm$. 

There are two vectors we can define: 
\begin{equation}\label{eq:vw}
	v_m = \eta^{2\,\dagger}_- \gamma_m \eta^1_+ \ ,\qquad
	w_m = \overline{\eta^2_-} \gamma_m \eta^1_+\ .
\end{equation}
In bispinor language, we can compute 
\begin{subequations} \label{eq:oddf}
\begin{align}
\eta_+ \eta_-^\dagger = \frac{1}{4}(1+\gamma) v\ , &\qquad \eta_+^c \eta_-^{c\, \dagger} = \frac{1}{4}(1+\gamma) \overline{v}\ , \\
\eta_- \eta_+^\dagger = \frac{1}{4}(1-\gamma) \overline{v}\ , &\qquad \eta_-^c \eta_+^{c\, \dagger} = \frac{1}{4}(1-\gamma) v\ ,
\end{align}
and 
\begin{align}
\eta_+ \eta_-^{c\, \dagger} = \frac{1}{4}(1+\gamma) w\ , &\qquad \eta_+^c \eta_-^\dagger = -\frac{1}{4}(1+\gamma) \overline{w}\ , \\
\eta_-\eta_+^{c\, \dagger} = -\frac{1}{4}(1-\gamma) w\ , &\qquad \eta_-^c \eta_+^\dagger = \frac{1}{4}(1-\gamma) \overline{w}\ .
\end{align}
\end{subequations}
(In four Euclidean dimensions, the chiral $\gamma = \ast_4 \lambda$, so that $(1+\gamma)v = v+ \ast_4v$, and so on. See \cite[App.~A]{10d} for more details.) 
For the more general case where $\eta^1_\pm \neq \eta^2_\pm$, we can simply refer back to (\ref{eq:eta12}). For example we get 
\begin{equation}\label{eq:eta+-}
\begin{split}
	\psi^1_+= \frac{e^{i(u_+-t_-)}}4 \cos(\alpha/2)\sin(\tilde \alpha/2)(1+\gamma) & \left[ \cos\left(\frac{\theta_+ + \theta_-}2\right) \Re v + i \cos\left(\frac{\theta_+ - \theta_-}2\right)  \Im v \; + \right . \\ &\left. - \sin\left(\frac{\theta_+ + \theta_-}2\right)  \Re w+ i \sin\left(\frac{\theta_+ - \theta_-}2\right) \Im w \right]\ .
\end{split}	
\end{equation}
For the time being we do not show the lengthy expressions for the other odd bispinors $\psi^2_+$ and $\psi^{1,2}_-$, because they will all turn out to simplify quite a bit as soon as we impose the zero-form equations in (\ref{eq:64}).

The $v$ and $w$ we just introduced are a complex vielbein; let us see why. First, a standard Fierz computation gives
\begin{equation}\label{eq:veta}
	v \cdot \eta_+ = 0 \ ,\qquad \overline v \cdot\eta_+  = 2 \eta_- \ ,
\end{equation}
where $\cdot$ denotes Clifford product. Multiplying from the left by $\eta^\dagger_-$, we obtain 
\begin{equation}\label{eq:v.v}
	v^2 = 0 \ ,\qquad v \, \llcorner \, \overline v = v^m \overline v_m  = 2 \ .
\end{equation}
Similarly to (\ref{eq:veta}), we can compute the action of $w$:
\begin{equation}\label{eq:weta}
	w \cdot \eta_\pm = 0 	\ ,\qquad \overline w \cdot \eta_\pm = \pm 2 \eta^c_\mp\ . 
\end{equation}
Multiplying by $\overline{\eta_\mp}$, we get
\begin{equation}\label{eq:w.w}
	w^2= 0 \ ,\qquad w \, \llcorner \, \overline w = 2\ .
\end{equation}
From (\ref{eq:veta}) we can also get $v \cdot \eta_+ \overline{\eta_-}= 0$, $\overline v \cdot \eta_+ \overline{\eta_-}= 2 \eta_- \overline{\eta_-}$, whose zero-form parts read
\begin{equation}\label{eq:v.w}
	v \, \llcorner \, w = 0 = \overline v \, \llcorner \, w  \ .
\end{equation}
Together, (\ref{eq:v.v}), (\ref{eq:w.w}), (\ref{eq:v.w}) say that
\begin{equation}\label{eq:viel}
	\{ {\rm Re} v,\ {\rm Re} w,\ {\rm Im} v,\ {\rm Im} w \}
\end{equation}
are a vielbein. 

We can also now try to relate the even forms of section \ref{sub:even} to this vielbein. From (\ref{eq:veta}) we also see $v\cdot \eta_+ \overline{\eta_+}=0$, which says $v\wedge \omega_+=0$; similarly one gets $\overline v \wedge \omega_-=0$. 
Also, (\ref{eq:weta}) implies that $w \cdot  \eta_+ \overline{\eta_+} = w \cdot \omega_+=0$, and thus that $w\wedge \omega_\pm=0$. So we have $\omega_+ \propto v \wedge w$, $\omega_- \propto \overline v \wedge w$. One can fix the proportionality constant by a little more work: 
\begin{subequations}\label{eq:relform}
\begin{equation}
	\omega_+ = - v \wedge w\ , \qquad \omega_- = \overline{v} \wedge w\ .
\end{equation}
Similar considerations also determine the real two-forms: 
\begin{equation}\label{eq:jpm}
	j_\pm = \pm\frac{i}{2}(v \wedge \overline{v} \pm w \wedge \overline{w})\ .
\end{equation}
\end{subequations}

So far we have managed to parameterize all the pure spinors $\Phi_\pm$, $\Psi_\pm$ in terms of a vielbein given by (\ref{eq:viel}). The expressions for $\Phi_+$ are given in (\ref{eq:eta++});  $\Phi_-$ is given by changing $(\cos(\alpha/2),\cos(\tilde \alpha/2))\to (\sin(\alpha/2),\sin(\tilde \alpha/2))$, and ${}_+ \to {}_-$ everywhere. The forms $j_\pm$, $\omega_\pm$ are given in (\ref{eq:relform}) in terms of the vielbein. Among the odd forms of $\Psi_\pm$, we have only quoted one example, (\ref{eq:eta+-}); similar expressions exist for $\psi^2_+$ and for $\psi^{1,2}_-$. We will summarize all this again after the simplest supersymmetry equations will allow us to simplify the parameterization quite a bit.



\section{General analysis} 
\label{sec:gen}

We will now use the parameterization obtained for $\Phi$ and $\Psi$ in section \ref{sec:para} in the system (\ref{eq:64}). As anticipated in the introduction, we will reduce the system to the two PDEs (\ref{eq:ddz}), (\ref{eq:a2}), and we will determine the local form of the metric and of the fluxes in terms of a solution to those equations.

\subsection{Zero-form equations} 
\label{sub:0}

The only equations in (\ref{eq:64}) that have a zero-form part are (\ref{eq:a}) and (\ref{eq:efg}): 
\begin{equation}\label{eq:0}
	(\Phi_+ + \Phi_-)^0_0 = 0 \ ,\qquad (\Phi_+ + \Phi_-)^\alpha_0 = 0 \ . 
\end{equation}
The subscript ${}_0$ here denotes the zero-form part. (Recall that the superscripts ${}^0$ and ${}^\alpha$ denote SU(2) singlets and triplets respectively.) To simplify the analysis, it is useful to change variables so as to make the SU(2) R-symmetry more manifest; this will lead us to definitions similar to those made in \cite[Sec.~4.5]{afrt}. 

In (\ref{eq:eta++}), apart for the overall factor $\cos(\alpha/2)\cos(\tilde \alpha/2)/4$, we have $\phi^1_{+\,0} \propto e^{i(u_+-t_+)} \cos(\theta_+)$, $\phi^2_{+\,0} \propto e^{i(u_++t_+)} \sin(\theta_+)$. The singlet is ${\rm Re} \phi^1_{+\,0} \propto \cos(\theta_+) \cos(u_+-t_+)$, and it is a good idea to give it a name, say $x_+$. On the other hand, the triplet is $\{{\rm Im}  \phi^2_+, {\rm Re} \phi^2_+, {\rm Im} \phi^1_+\}  \propto \{\sin(\theta_+) \sin(u_+ + t_+) , \sin(\theta_+) \cos(u_+ + t_+), \cos(\theta_+) \sin(u_+ - t_+)\}$. If we sum their squares, we obtain:
\begin{equation}\label{eq:trsq}
\sin^2(\theta_+) + \cos(\theta_+)^2 \sin^2(u_+-t_+) = x_+^2 \tan^2(u_+-t_+) + \sin^2(\theta_+)= 1-x_+^2\ .
\end{equation}
This suggests that we parameterize the triplet using the combination $\sqrt{1-x^2_+}\, y^\alpha$, where $y^\alpha$ should obey $y_\alpha y^\alpha =1$ and can be chosen to be the $\ell=1$ spherical harmonics on $S^2$. What we are doing is essentially changing variables on an $S^3$, going from coordinates that exhibit it as an $S^1\times S^1$ fibration over an interval to coordinates that exhibit it as an $S^2$ fibration over an interval:
\begin{equation}
	\left\{ \cos(\theta_+) e^{i(u_+-t_+)}, \sin(\theta_+) e^{i(u_++t_+)}\right\} \to 
	\left\{ x_+, \sqrt{1-x_+^2} y^\alpha \right\}\ .
\end{equation}
An identical discussion can of course be given for $\phi^{1,2}_-$. Summing up,  we are led to the following definitions:
\begin{equation}\label{eq:xbg}
	x_\pm \equiv \cos(\theta_\pm) \cos(u_\pm - t_\pm) \ ,\quad
	\sin\beta_\pm \equiv \frac{\sin(\theta_+)}{\sqrt{1-x_+^2}} \ ,\quad
	\gamma_\pm \equiv \frac\pi2 - u_\pm - t_\pm \ , 
\end{equation}
and 
\begin{equation}\label{eq:y}
	y^\alpha_\pm \equiv \Big(\sin(\beta_\pm) \cos(\gamma_\pm), \ \sin(\beta_\pm) \sin(\gamma_\pm), \ \cos(\beta_\pm)\Big) \ ,
\end{equation}
in terms of which 
\begin{equation}\label{eq:Phipm0}
\begin{split}
	\Phi_{+\,0}= \cos(\alpha/2) \cos(\tilde\alpha/2) \left(x_+ + i y^\alpha_+ \sqrt{1-x^2_+}\sigma_\alpha\right) 
	\ ,\\
	\Phi_{-\,0}= \sin(\alpha/2) \sin(\tilde\alpha/2) \left(x_- + i y^\alpha_- \sqrt{1-x^2_-}\sigma_\alpha\right) \ .	
\end{split}
\end{equation}

Going back to (\ref{eq:0}), summing the squares of all four equations we get $\cos^2(\alpha/2) \cos^2(\tilde \alpha/2) = \sin^2(\alpha/2) \sin^2(\tilde \alpha/2)$. Given that $\alpha$ and $\tilde \alpha \in [0,\pi]$, this is uniquely solved by 
\begin{equation}
	\tilde \alpha = \pi - \alpha \ .
\end{equation}
Now (\ref{eq:0}) reduces to 
\begin{equation}\label{eq:x+-}
	-x_- = x_+ \equiv x \ ,\qquad -y^\alpha_- = y^\alpha_+ \equiv y^\alpha \ .
\end{equation}
In terms of the original parameters, this means $\theta_+ = \theta_- $, $u_- = u_+$, $t_- = t_+ + \pi$. 

The parameterization obtained in section \ref{sec:para} now simplifies considerably: 
\begin{subequations}\label{eq:paras}
\begin{align}
&\phi^1_\pm = \pm\frac18 \sin\alpha \cos\theta\, e^{i(u-t)} \exp\left[-\frac1{\cos\theta}(i j_\pm + \sin\theta {\rm Re} \omega_\pm)\right] \ , \\
&\phi^2_\pm = \pm\frac18 \sin\alpha \sin\theta\, e^{i(u+t)} \exp\left[\frac1{\sin\theta}(\cos\theta {\rm Re} \omega_+ + i {\rm Im}  \omega_+)\right]\ ; \\
&\psi^1_\pm = \mp \frac18 (1\pm \cos\alpha )e^{i(u-t)} (1 \pm \gamma) \left[ \cos\theta {\rm Re} v \pm i {\rm Im} v \mp \sin\theta {\rm Re} w\right] \ , \label{eq:paraspsi1} \\ 
&\psi^2_\pm = \mp \frac18 (1\pm \cos\alpha )e^{i(u+t)} (1 \pm \gamma) \left[ \sin\theta {\rm Re} v \pm i {\rm Im} w \pm \cos\theta {\rm Re} w\right] \ . \label{eq:paraspsi2}
\end{align}	
\end{subequations}
We temporarily reverted here to a formulation where SU$(2)_{\rm R}$ is not manifest; however, in what follows we will almost always use the SU$(2)$-covariant variables $x$ and $y^\alpha$ introduced above. 


\subsection{Geometry} 
\label{sub:geo}

We will now describe how we analyzed the higher-form parts of (\ref{eq:64}), although not in such detail as in section \ref{sub:0}. 

The only equations that have a one-form part are (\ref{eq:bcd}). From (\ref{eq:paraspsi1}), (\ref{eq:paraspsi2}), we see that the second summand $(\Psi_+ + \Psi_-)^\alpha_1$ is a linear combination of the forms in the vielbein (\ref{eq:viel}). The first summand consists of derivatives of the parameters we have previously introduced. This gives three constraints on the four elements of the vielbein. We used it to express ${\rm Im} v$, ${\rm Re} w$, ${\rm Im} w$ in terms of ${\rm Re} v$;\footnote{Doing so requires $x\neq 0$; the case $x=0$ will be analyzed separately in section \ref{sub:tBO}.} the resulting expressions are at this point still not particularly illuminating, and we will not give them here. These expressions are not even manifestly SU$(2)$-covariant at this point; however, once one uses them into $\Phi_\pm$ and $\Psi_\pm$, one does find SU$(2)$-covariant forms. Just by way of example, we have 
\begin{equation}
\begin{split}
	&(\Phi_+ + \Phi_-)^\alpha_2 = -\frac13 e^{-3A+\phi} \sin\alpha \, {\rm Re} v \wedge 
	d\left( y^\alpha \sin\alpha \, e^{4A-\phi} \sqrt{1-x^2}\right) \ , \\
	&(\Psi_- - \Psi_+)^\alpha_1 = y^\alpha \sqrt{1-x^2} \sin^2(\alpha){\rm Re} v + \frac13 e^{-3A+\phi}\cos\alpha  \,d\left( y^\alpha \sin\alpha  \, e^{4A-\phi} \sqrt{1-x^2}\right)\ .
\end{split}
\end{equation}
We chose these particular 2-form and 1-form triplet combinations because they are involved in the 2-form part of (\ref{eq:efg}). The result is a triplet of equations of the form $y^\alpha\, E_2 + dy^\alpha \wedge E_1=0$, where $E_i$ are $i$-forms and SU(2)$_{\rm R}$ singlets. If we multiply this by $y_\alpha$, we obtain $E_2=0$ (since $y_\alpha dy^\alpha = 0$); then also $E_1=0$ necessarily. The latter gives a simple expression for ${\rm Re} v$, the one-form among the vielbein (\ref{eq:viel}) that we had not determined yet:
\begin{equation} \label{eq:rev}
	\Re v = -\frac{e^{-A}}{\sin\alpha }d(e^{2A}\cos\alpha )\ .
\end{equation}
Once this is used, the two-form equation $E_2=0$ is automatically satisfied.

There are some more two-form equations from (\ref{eq:64}). The easiest is (\ref{eq:i}), which gives
\begin{subequations}
\begin{equation}\label{eq:ddz}
	d\left( \frac{e^{4A-\phi}}x \cot\alpha\,  d(e^{2A} \cos\alpha )+\frac1{3x} e^{2A}\sqrt{1-x^2} d\left(e^{4A-\phi}\sqrt{1-x^2}\sin\alpha \right)\right)=0\ .
\end{equation}
Locally, this can be solved by saying 
\begin{equation}\label{eq:dz}
	\boxed{x dz= e^{4A-\phi} \cot\alpha\,  d(e^{2A} \cos\alpha )+\frac13 e^{2A}\sqrt{1-x^2} d\left(e^{4A-\phi}\sqrt{1-x^2}\sin\alpha \right)}
\end{equation}
\end{subequations}
for some function $z$. The two-form part of (\ref{eq:a}) reads, on the other hand, 
\begin{equation}\label{eq:a2}
	\boxed{e^{-8A}d(e^{6A}\cos\alpha )\wedge dz = d(x e^{2A-\phi}\sin\alpha ) \wedge d(e^{2A}\cos\alpha ) }\ .
\end{equation}
If one prefers, $dz$ can be eliminated, giving
\begin{equation}\label{eq:a2noz}
	3 \sin(2 \alpha) dA\wedge d \phi =  d \alpha\wedge \Big(6dA+\sin^2(\alpha)\left(-dx^2 -2(x^2+5) dA + (1+2x^2)d \phi\right)\Big)\ .
\end{equation}
We will devote the whole section \ref{sec:pde} to analyze the PDEs (\ref{eq:ddz}), (\ref{eq:a2}) and we will also exhibit two explicit solutions.

Taking the exterior derivative of (\ref{eq:a2}) one sees that $d \alpha\wedge dA \wedge dz=0$. Wedging (\ref{eq:ddz}) with an appropriate one-form, one also sees $d \alpha \wedge dA \wedge dx=0$. Taken together, these mean that only two among the remaining variables $(\alpha,x,A,\phi)$ are really independent. For example we can take $\alpha$ and $x$ to be independent, and      
\begin{equation}\label{eq:dep}
	A=A(\alpha,x) \ ,\qquad \phi= \phi(\alpha,x)\ .
\end{equation}

We are not done with the analysis of (\ref{eq:64}), but there will be no longer any purely  geometrical equations: the remaining content of (\ref{eq:64}) determines the fluxes, as we will see in the next subsection. Let us then pause to notice that at this point we have already determined the metric: three of the elements of the vielbein (\ref{eq:viel}) were determined already at the beginning of this section in terms of $\Re v$, and the latter was determined in (\ref{eq:rev}). 
%
This gives the metric
\begin{equation}\label{eq:metpq}
	ds^2 = \frac{\cos\alpha}{\sin^2(\alpha)}\frac{dq^2}{q} + \frac19 q(1-x^2)\frac{\sin^2(\alpha)}{\cos\alpha }
	\left( \frac1{x^2}\left(\frac{dp}p +3\cot^2(\alpha)\frac{dq}q\right)^2 + ds^2_{S^2}\right)\ ,
\end{equation}
where the $S^2$ is spanned by the functions $\beta$ and $\gamma$ introduced in (\ref{eq:y}) (namely, $ds^2_{S^2}= d \beta^2 + \sin^2(\beta) d \gamma^2$), and we have eliminated $A$ and $\phi$ in favor of 
\begin{equation}
	q \equiv e^{2A}\cos\alpha  \ ,\qquad p \equiv e^{4A-\phi}\sin\alpha \sqrt{1-x^2}\ . 
\end{equation}
These variables could also be used in the equations (\ref{eq:ddz}), (\ref{eq:a2}) above, with marginal simplification. Notice that positivity of (\ref{eq:metpq}) requires $|x|\le 1$.

Thus we have found in this section that the internal space $M_4$ is an $S^2$ fibration over a two-dimensional space $\Sigma$, which we can think of as spanned by the coordinates $(\alpha,x)$. 
 

\subsection{Fluxes} 
\label{sub:flux}

We now turn to the three-form part of (\ref{eq:bcd}). This is an SU(2)$_{\rm R}$ triplet. It can be written as $y^\alpha H = \epsilon^{\alpha \beta \gamma} y^\beta d y^\gamma \wedge \tilde E_2 + y^\alpha \vol_{S^2} \wedge \tilde E_1$, where $\tilde E_i$ are $i$-forms and SU(2)$_{\rm R}$ singlets. Actually, from (\ref{eq:ddz}) and (\ref{eq:a2}) it follows that $\tilde E_2=0$; we are then left with a single equation setting $H = \vol_{S^2} \wedge \tilde E_1$:
\begin{equation}\label{eq:H}
	H = -\frac1{9x} e^{2A} \sqrt{1-x^2} \sin\alpha  \left[ -\frac{6dA}{\sin\alpha } + 2 e^{-A}(1+x^2)d(e^A\sin\alpha )+ \sin\alpha\,  d(\phi + x^2)\right]\wedge \vol_{S^2}\ .
\end{equation}
As expected, $H$ is a singlet under SU(2)$_{\rm R}$. 

All the four-form equations in (\ref{eq:i}), (\ref{eq:a}), (\ref{eq:efg}) turn out to be automatically satisfied. We can then finally turn our attention to (\ref{eq:h}), which we have ignored so far. It gives the following expressions for the fluxes:
\begin{subequations}\label{eq:RR}
\begin{equation}\label{eq:f1} 
F_1 = \frac{e^{-\phi}}{6x\cos\alpha }\left[ \frac{12dA}{\sin\alpha } + 4 e^{-A}(x^2-1)d(e^A \sin\alpha ) + e^{2 \phi}\sin\alpha \, d(e^{-2 \phi}(1+2x^2))\right]\ ;
\end{equation}
\begin{align}\label{eq:f3} 
F_3 = \frac{e^{2A-\phi}}{54} \sqrt{1-x^2} \frac{\sin^2(\alpha)}{\cos\alpha } &\left[ \frac{36dA}{\sin\alpha }+4 e^{-A} (x^2-7) d(e^A \sin\alpha ) \; + \right. \nonumber \\ & \big. + \ e^{2 \phi} \sin\alpha\,  d(e^{-2 \phi} (1+2x^2))\bigg] \wedge \vol_{S^2}\ .	
\end{align}
\end{subequations}

The Bianchi identities 
\begin{equation}
	d H = 0 \ ,\quad d F_1 = 0 \ ,\quad d F_3 + H \wedge F_1 = 0 \ ,
\end{equation}
are all automatically satisfied, using of course the PDEs (\ref{eq:ddz}), (\ref{eq:a2}). As usual, this statement is actually true only if one assumes that the various functions appearing in those equations are smooth. As in \cite{afrt}, one can introduce sources by relaxing this condition. 


\subsection{The case $x=0$} 
\label{sub:tBO}

In section \ref{sub:geo}, we used the three-form part of (\ref{eq:bcd}) to express ${\rm Im} v$, ${\rm Re} w$, ${\rm Im} w$ in terms of ${\rm Re} v$. This actually can only be done for $x\neq 0$: the expressions we get contain $x$ in the denominator, as can be seen for example in (\ref{eq:ddz}). This left out the case $x=0$; we will analyze it in this section, showing that it leads to a single solution, discussed in \cite{cvetic-lu-pope-vazquezporitz,lozano-colgain-rodriguezgomez-sfetsos} --- namely, to a T-dual of the AdS$_6$ solution found in \cite{brandhuber-oz} and reviewed in our language in appendix \ref{app:IIA}. 

Keeping in mind that $-x_- = x_+ = x$ (from (\ref{eq:x+-})), from (\ref{eq:xbg}) we have $x=\cos(\theta) \cos(u-t)$. Imposing $x=0$ then means either $\theta= \frac \pi 2$ or $u-t=\frac \pi2 $. Of these two possibilities, the first does not look promising, because on the $S^3$ parameterized by $(\cos(\theta)e^{i(u-t)}, \sin(\theta)e^{i(u+t)})$ it effectively restricts us to an $S^1$: only the function $u+t$ is left in the game, and indeed going further in the analysis one finds that the metric becomes degenerate.\footnote{At the stage of (\ref{eq:vi3x0}) below, one would find ${\rm Re} w \propto {\rm Im} v$.} The second possibility, $u-t=\frac \pi2 $, restricts us instead to an $S^2\subset S^3$; we will now see that this possibility survives. It gives 
\begin{equation}
	\beta = \theta \ ,\qquad t= -\frac12 \gamma \ ,\qquad u = \frac \pi 2 - \frac12 \gamma \ .
\end{equation}
This leads to a dramatic simplification in the whole system. The one-form equations from (\ref{eq:bcd}) do not involve ${\rm Im} v$ any more; we can now use them to solve for ${\rm Re} v$, ${\rm Re} w$, ${\rm Im} w$ (rather than for ${\rm Im} v$, ${\rm Re} w$, ${\rm Im} w$ as we did in previous subsections, for $x\neq 0$). This strategy would actually have been possible for $x\neq 0$ too, but it would have led to far more involved expressions; for this reason we decided to isolate the $x=0$ case and to treat it separately in this subsection. We get
\begin{equation}\label{eq:vi3x0}
	{\rm Re} v= \frac{e^{-3A+ \phi}}{3\cos\alpha } d(\sin\alpha e^{4A-\phi}) \ ,\quad
	{\rm Re} w = \frac{e^A}3 \sin\alpha  \,d \beta \ ,\quad
	{\rm Im} w = -\frac{e^A}3 \sin\alpha \, \sin\beta d \gamma \ .
\end{equation}

We now turn to the 2-form equation in (\ref{eq:efg}). As in the previous subsections of this section, this can be separated into a 2-form multiplying $y^\alpha$ and a 1-form multiplying $d y^\alpha$, which have to vanish separately: 
\begin{equation}\label{eq:efgx0}
	d(e^{5A-\phi} {\rm Re} v) = 0 \ ,\qquad
	e^{5A-\phi} (3-4 \sin^2(\alpha)) {\rm Re} v  = d(e^{6A-\phi}\sin\alpha \cos\alpha )\ .
\end{equation} 
Hitting the second equation with $d$ and using the first, we find $\sin\alpha \cos\alpha \,d \alpha \wedge {\rm Re} v=0$, and hence, recalling (\ref{eq:vi3x0}), to $\sin\alpha  d \alpha\wedge d(4A-\phi)=0$. Now, $\sin\alpha $ is not allowed to vanish because of (\ref{eq:vi3x0}) (recall that ${\rm Re} v$, ${\rm Re} w$, ${\rm Im} w$ are part of a vielbein); hence $d \alpha\wedge d(4A-\phi)=0$. This can be interpreted as saying that $4A-\phi$ is a function of $\alpha$. On the other hand, using (\ref{eq:vi3x0}) in the first in (\ref{eq:efgx0}), we get $d(\frac{e^{2A}}{\cos\alpha })\wedge d(\sin\alpha  e^{4A-\phi})=0$, which shows that $A=A(\alpha)$, and hence also that $\phi=\phi(\alpha)$. Going back to the second in (\ref{eq:efgx0}), it now reads
\begin{equation}\label{eq:odeefg2x0}
	2(\cos^2(\alpha)+2) \del_\alpha A + \sin^2(\alpha) \del_\alpha \phi =  \sin(2\alpha)\ .
\end{equation}

Turning to (\ref{eq:i}), its 2-form part reads
\begin{equation}\label{eq:ix0}
	d(e^{5A-\phi} \Im v)=0 \qquad \Rightarrow \qquad {\rm Im}  v = e^{-(5A-\phi)} dz \ 
\end{equation}
for some function $z$. This completes (\ref{eq:vi3x0}). 

Finally, (\ref{eq:a}) gives 
\begin{equation}\label{eq:ax0}
	\left(d(e^{-2A}\cos\alpha ) + 2 e^{-3A} \sin\alpha  {\rm Re} v\right) \wedge {\rm Im} v = 0 \ .
\end{equation}
In view of (\ref{eq:ix0}), the parenthesis has to vanish by itself; this leads to 
\begin{equation}\label{eq:odeax0}
	4(7 \cos^2(\alpha)-4) \del_\alpha A + 4\sin^2(\alpha)\del_\alpha \phi = -\sin(2\alpha)\ .
\end{equation}

Notice that now (\ref{eq:odeefg2x0}) and (\ref{eq:odeax0}) are two \emph{ordinary} (as opposed to partial) differential equations, which can be solved explicitly:
\begin{equation}\label{eq:abTsol}
	e^A= \frac{c_1}{\cos^{1/6}(\alpha)} \ ,\qquad e^\phi = \frac{c_2}{\sin\alpha  \cos^{2/3}(\alpha)}\ ,
\end{equation}
where $c_i$ are two integration constants. These are exactly the warping and dilaton presented in \cite[(A.1)]{lozano-colgain-rodriguezgomez}, for $c_1=\frac32 L m^{-1/6}$, $c_2 =4/(3L^2 m^{2/3})$. It is now possible to derive the fluxes, as we did in subsection \ref{sub:flux} for $x\neq0$, and check that they coincide with those in \cite[(A.1)]{lozano-colgain-rodriguezgomez}. 

The metric can now be computed too, using the vielbein (\ref{eq:efgx0}), (\ref{eq:ix0}); it also agrees with the one given in \cite[(A.1)]{lozano-colgain-rodriguezgomez}. It inherits the singularity at $\alpha=\frac \pi2$ from the Brandhuber--Oz solution \cite{brandhuber-oz}; moreover, it now has a singularity at $\alpha=0$. The latter is actually the singularity one always gets when one T-dualizes along a Hopf direction in a $S^3$ that shrinks somewhere. It represents an NS5 smeared along the T-dual $S^1$; one expects worldsheet instantons to modify the metric so that the NS5 singularity gets localized along that direction, as in \cite{tong}. As for the singularity at $\alpha=\frac\pi2$, it now cannot be associated with an O8--D8 system as it was in IIA, since we are in IIB. It probably now represents a smeared O7--D7 system; it is indeed always the case that T-dualizing a brane along a parallel direction in supergravity gives a smeared version of the correct D-brane solution on the other side, as we just saw for the NS5-brane. It is possible that again instanton effects localize the singularity, this time to an O7--D7 system. (Even more correctly, we should expect the O7 to split into an $(1,1)$-sevenbrane and an $(1,-1)$-sevenbranes, as pointed out in \cite{dewolfe-hanany-iqbal-katz} following \cite{sen-O7}.)
 
Notice finally that, although we have found it convenient to treat the $x=0$ case separately from the rest, it is in fact a particular case of the general treatment (although a slightly degenerate one). Indeed one can check that (\ref{eq:dz}) is satisfied by (\ref{eq:abTsol}); in contrast to the general case, this does not determine a function $z$, but we can use (\ref{eq:a2noz}), where $z$ has been eliminated, instead of (\ref{eq:a2}), which contains $z$. Thus the solution presented in this subsection is already an example of our general formalism. In section \ref{sub:nonabt} we will see another, more elaborate example.



\section{The PDEs} 
\label{sec:pde}

In section \ref{sec:gen}, we reduced the problem of finding AdS$_6\times M_4$ solutions to the two PDEs (\ref{eq:ddz}), (\ref{eq:a2}). As anticipated in the introduction, we will not try to find the most general solution to these equations in this paper. In this section we will make some general remarks about the PDEs, and we will recover via a simple Ansatz the known solution \cite{lozano-colgain-rodriguezgomez-sfetsos}, originally obtained via nonabelian T-duality. (As we mentioned in that section, one can also see the $x=0$ case as a particular solution to the PDEs.)

\subsection{General considerations} 
\label{sub:pde}

We derived in section \ref{sub:geo} the two equations (\ref{eq:ddz}), (\ref{eq:a2}). Recall that $z$ is an auxiliary variable, defined by (\ref{eq:dz}). As we already remarked, among the four remaining variables $(\alpha,x,A,\phi)$, only two (for example $\alpha$ and $x$) are independent. The other two, $A$ and $\phi$, can be taken to be dependent as in (\ref{eq:dep}). The equations (\ref{eq:ddz}) and (\ref{eq:a2}) can then be reexpressed as two scalar PDEs in the two dimensions spanned by $\alpha$ and $x$:
\begin{subequations}
\begin{align}
	3\sin(2 \alpha) (A_\alpha \phi_x - A_x \phi_\alpha) &= 6 A_x + \sin^2 \alpha \left(-2x -2 (x^2+5) A_x+(1+2x^2)\phi_x\right)\ , \\
	\cos \alpha (2+3 x \phi_x)+\sin \alpha \phi_\alpha &= 2x \left(\frac 3{\sin \alpha}+(x^2-4)\sin \alpha\right)(A_\alpha \phi_x - A_x \phi_\alpha) \ + \\
	&-2x \cos \alpha\left(\frac 3{\sin^2\alpha}-(5+x^2)\right)A_x + 2\left(\frac 3{\sin \alpha}-(1+x^2)\sin \alpha\right) A_\alpha \ ,\nonumber 
\end{align}	
\end{subequations} 
where $A_\alpha \equiv \del_\alpha A$ etc. As we will see, they are actually easier to study in their original form manifestations (\ref{eq:ddz}) and (\ref{eq:a2}).

These equations are nonlinear, and as such they are rather hard to study. Even so, there are quite a few techniques that have been developed over the years to tackle such systems. Perhaps the first natural question is how many solutions one should expect. For a first-order system of ODEs, it is roughly enough to compare the number of equations to the number of functions. If there are $n$ equations and $n$ functions, the system is neither over- nor under-constrained: geometrically, the system gives a vector field in an open set in $\rr^{n+1}$ (including time), and solving the system means finding integral curves to this vector field. (When the system is ``autonomous'', i.e.~it does not depend explicitly on time, one can more simply consider a vector field on $\rr^n$). 

The picture is more complicated for a system of PDEs. In general, if we have $k$ ``times''  and $m$ functions, the system will define a distribution of dimension $k$ (namely, a choice of subspaces $V_x \subset T_x \rr^{k+m}$ of dimension $k$ for every point $x\in \rr^{k+m}$); solving the system then means finding ``integral submanifolds'' for the distribution, namely submanifolds $S \subset \rr^{k+m}$ such that $V_x$ is tangent to $S$ for every $x\in S$. This distribution is in general however not guaranteed to admit integral submanifolds. (A famous example is given by Frobenius theorem: a distribution defined by the span of vector fields $v_i$ will only be integrable if all the Lie brackets $[v_i,v_j]$ are linear combinations of the $v_i$ themselves.) Fortunately, the machinery of ``exterior differential systems'' (EDS) has been developed to deal with these issues, culminating in the Cartan--K\"ahler theorem (see for example \cite[Chap.III]{bryant-chern-gardner-goldshmidt-griffiths}, or \cite[Sec.~10.4.1]{stephani-kramer-maccallum-hoenselaers-herlt} in slightly more informal language).

Describing and applying such methods in detail is beyond the scope of this paper, but here is a sketch. First one defines a ``differential ideal'', namely a vector space of the equations in the system and their exterior derivatives. In our case, denote by $E_i$ the two two-forms that have to vanish in (\ref{eq:ddz}) and (\ref{eq:a2}); the ideal is then the linear span $I=\langle E_1, E_2, dE_2 \rangle$ (since $dE_1=0$ automatically). 
We then want to construct the distribution $V$ on which the forms in $I$ vanish, in the sense that each multi-vector built from vectors in the distribution has zero pairing with the forms in $I$.
One proceeds iteratively. We first consider a single vector field $e_1$ on which the forms vanish (in our case this is trivial, since there are no one-forms in $I$; we can take for example $e_1=\del_\alpha$). We then add a second vector: this is done by solving the ``polar equations'' $H(E_1)\equiv\{v\llcorner e_1 \llcorner E_i=0\}$. The rank of this system is  denoted by $c_1$. (In general there might be a $c_0$ too, but in our case the first choice of a vector was free because there are no one-forms in $I$; $c_0$ is then considered to be $0$.) For us it turns out that $c_1=2$. In general one would go on by choosing a solution $e_2$ to the polar equations above, and would consider new polar equations $H(E_2)\equiv\{v\llcorner e_1 \llcorner E_i=v\llcorner e_2 \llcorner E_i=0, v\llcorner e_1 \llcorner e_2 \llcorner dE_2=0\};$ the rank of this new system would be denoted by $c_2$, which in our case also happens to be $2$. However, solving our PDEs means finding a two-dimensional integral manifold, and hence we can stop at the second step and disregard the higher polar equations $H(E_2)$. (The general theory would also show that for our system there is actually no three-dimensional integral manifold.) We can then apply the so-called ``Cartan test'' and a corollary to the Cartan--K\"ahler theorem (respectively Thm.~1.11 and Cor.~2.3 in \cite{bryant-chern-gardner-goldshmidt-griffiths}) to infer that an integral submanifold of dimension 2 actually does exist. The proof of the theorem also says that the general solution depends on $s_1=c_1-c_0=2$ functions of one variable. ($s_i=c_i-c_{i-1}$ are called ``Cartan characters''.) These two functions can be thought of as functions at the boundary of the two-dimensional domain in $\alpha$ and $x$ on which the solution exists.

Having determined the structure of the solutions, it would be nice to find as many as possible of them. A strategy which is common in this context is to impose some extra symmetry. This is less obvious than usual to implement. We cannot for example just assume that $A$ and $\phi$ do not depend on one of the coordinates $\alpha$ and $x$: the metric (\ref{eq:metpq}) would become degenerate. Another perhaps more promising idea is to use the so-called ``method of characteristics'' to reduce the problem to a system of ODEs. We plan to return on this in the future.

Finally, let us point out that two solutions to our PDEs are already known. One is the case $x=0$, which we studied in section \ref{sub:tBO}. Although we had to treat it separately, we also mentioned that it is a solution of the general system of PDEs (once we eliminate $dz$ from (\ref{eq:a2}), obtaining (\ref{eq:a2noz})). 

We will now see another particular solution. Although the global properties of the resulting $M_4$ are even more puzzling than those of the solution in section \ref{sub:tBO}, it might be possible to generalize it to new solutions which are better-behaved; for example, one might start by studying perturbations around it.


\subsection{A local solution: nonabelian T-duality} 
\label{sub:nonabt}

Many PDEs are reduced to ODEs by a separation of variables Ansatz. For our nonlinear PDEs, this does not work. However, we will now see that a particular case does lead to a solution, namely:
\begin{equation}\label{eq:nonabAn}
	\phi= f(\alpha) + \log(x) \ ,\qquad A=A(\alpha)\ .
\end{equation}
Notice that this Ansatz restricts $x$ to be in $(0,1]$. (We already observed after (\ref{eq:metpq}) that $|x|\le 1$ in general.)

We begin by considering (\ref{eq:dz}). With (\ref{eq:nonabAn}), after a few manipulations it reduces to 
\begin{equation}
\begin{split}\label{eq:dznonab}
	dz=\; & d\left(e^{6A-f}\frac{\sin\alpha }{6x^2}\right) -\frac13 e^{2A}d(e^{4A-f}\sin\alpha )\; + \\
	& +\frac1{x^2}\left[	-\frac16 e^{4A} d(e^{2A-f}\sin\alpha )+e^{4A-f}\cot\alpha\, d(e^{2A}\cos\alpha )\right] \ .
\end{split}	 
\end{equation}
The first line in (\ref{eq:dznonab}) is manifestly exact, since everything is a function of $\alpha$ alone. The second line is of the form $\frac1{x^2} d({\rm function}(\alpha))$, and cannot be exact unless it vanishes, which leads to
\begin{equation}\label{eq:2line}
	d(e^{2A-f} \sin\alpha ) = 6 e^{-f} \cot\alpha \, d(e^{2A} \cos\alpha ) \ .
\end{equation}
The first line in (\ref{eq:dznonab}) then determines $dz$ (and can be integrated to produce $z$). We can now use this expression for $dz$ in (\ref{eq:a2}). Most terms in (\ref{eq:a2}) actually vanish because they involve wedges of forms proportional to $d \alpha$; the only one surviving is of the form $d(e^{6A}\cos\alpha )\wedge dx$. In other words, we are forced to take 
\begin{equation}
	e^A = c_1 (\cos\alpha)^{-1/6}\ ,
\end{equation}
with $c_1$ an integration constant. Plugging this back into (\ref{eq:2line}) we get 
\begin{equation}
	e^f  = c_2 \frac{(\cos\alpha)^{-1/3}}{\sin^3\alpha} 
\end{equation}
for $c_2$ another integration constant. 

This is actually the solution found in \cite{lozano-colgain-rodriguezgomez-sfetsos}. To see this, one needs to identify
\begin{equation}
	\alpha= \theta \ ,\qquad	x= \frac{e^{2\hat A}}{\sqrt{r^2 + e^{4\hat A}}}\ ,
\end{equation}
where $\hat A$ is the function denoted by $A$ in \cite{lozano-colgain-rodriguezgomez-sfetsos}. One can check that indeed the fluxes (\ref{eq:H}), (\ref{eq:RR}) and metric (\ref{eq:metpq}) give the expressions in \cite{lozano-colgain-rodriguezgomez-sfetsos}. 
The metric one gets has a singularity at $\alpha=\pi/2$, just like the solution \cite{brandhuber-oz}, and a new singularity at $\alpha=0$ \cite{lozano-colgain-rodriguezgomez}. More worryingly, it is noncompact; it might be possible to find a suitable analytic continuation, with the help of the PDEs (\ref{eq:ddz}), (\ref{eq:a2}) found in this paper.



\section*{Acknowledgments}
We would like to thank D.~Rodr\'iguez G\'omez, N. Kim, D.~Martelli, E.~{\'O Colg\'ain}, D.~Thompson and A.~Zaffaroni for interesting discussions. F.A.~is grateful to the Graduiertenkolleg GRK 1463 ``Analysis, Geometry and String Theory'' for support. The work of M.F.~was partially supported by the ERC Advanced Grant ``SyDuGraM'', by IISN-Belgium (convention 4.4514.08) and by the ``Communaut\'e Fran\c{c}aise de Belgique" through the ARC program. M.F.~ is a Research Fellow of the Belgian FNRS-FRS. A.P., D.R.~and A.T.~are supported in part by INFN, by the MIUR-FIRB grant RBFR10QS5J ``String Theory and Fundamental Interactions'', and by the MIUR-PRIN contract 2009-KHZKRX. The research of A.P. and A.T.~is also supported by the European Research Council under the European Union's Seventh Framework Program (FP/2007-2013) -- ERC Grant Agreement n. 307286 (XD-STRING). 

\appendix

\section{Derivation of (\ref{eq:64})} 
\label{app:10-6}

The starting point to obtain \eqref{eq:64} is the system of equations (3.11) in \cite{10d}, which was shown in that reference to be equivalent to $\mathcal{N}=1$ supersymmetry on any $M_{10}$; here we will specialize the ten-dimensional spacetime $M_{10}$ to $\ads$. Actually, the equations appearing in \eqref{eq:64} strictly derive from equations (3.1a) and (3.1b) in \cite{10d}. In section \ref{appsub:master} we show such derivation. 

Furthermore, to prove the equivalence between the system \eqref{eq:64} and the conditions imposed by $\mathcal{N}=1$ supersymmetry on $\ads$, we need to show that the two remaining ``pairing'' equations (3.1c,d) in \cite{10d} are completely redundant on such background: this is done in subsection \ref{appsub:pair}.

\subsection{Derivation of the system} 
\label{appsub:master}

Let us quote equations (3.1a,b) of \cite{10d}:
\begin{subequations}
\begin{align}
& d_H (e^{-\phi} \Phi) = -(\tilde{K} \wedge + \iota_K) F_{(10)}\ ; \label{eq:master} \\
&L_K g = 0\ , \quad d\tilde{K} = \iota_K H\ . \label{eq:symm10d}
\end{align}
\end{subequations}
$\Phi = \epsilon_1 \otimes \overline{\epsilon_2}$ is the key ten-dimensional polyform,\footnote{It should not be confused with the SU$(2)$-covariant internal even forms $\Phi_\pm$.} which is adapted to our background; $g$ is the ten-dimensional metric while $K$ and $\tilde{K}$ are ten-dimensional one-forms which will be defined momentarily.

The decomposition of the ten-dimensional spinors $\epsilon_a$ suggests we decompose accordingly the ten-dimensional gamma matrices:
\begin{equation}\label{eq:gamma64}
	\gamma^{(6+4)}_\mu = e^A \gamma^{(6)}_\mu \otimes 1 \ ,\qquad
	\gamma^{(6+4)}_{m+5} = \gamma^{(6)} \otimes \gamma^{(4)}_m \ .
\end{equation}
Here $\gamma^{(6)}_\mu$, $\mu=0,\ldots,5$, are a basis of six-dimensional gamma matrices ($\gamma^{(6)}$ is the chiral gamma), while $\gamma^{(4)}_m$, $m=1,\ldots,4$ are a basis of four-dimensional gamma matrices. We can now expand via Fierz identities (see formula (A.12) in \cite{10d}) the bilinear $\epsilon_1 \otimes \overline{\epsilon_2}$, by plugging in the decomposition \eqref{eq:10deps} and \eqref{eq:gamma64}. We get a sum of terms such as the following:
\begin{equation}\label{eq:fierz}
\sum_{k=0}^6 \frac{1}{8k!} \left(\overline{\zeta_+}  \gamma^j_{(6)} \gamma^{(6)}_{\mu_k \ldots \mu_1} \zeta_+ \right) \gamma_{(6)}^{\mu_1 \ldots \mu_k} \sum_{j=0}^4 \frac{1}{4j!} \left( \eta^{2\dagger}_\mp \gamma^{(4)}_{m_j \ldots m_1} \eta^1_+\right) \gamma_{(4)}^{m_1 \ldots m_j} = \mp \zeta_+ \overline{\zeta_+} \wedge \eta^1_+ \eta^{2\dagger}_\mp\ .
\end{equation}
What we mean by e.g. $\zeta_+ \overline{\zeta_+}$ is the six-dimensional polyform corresponding to this bilinear via the Clifford map (see footnote \ref{foot:cliff}). All in all we get:
\begin{equation}
\begin{split}
	\label{eq:Phispin}
	\Phi = & \ \mp \zeta_+ \overline{\zeta_+} \wedge \eta^1_+ \eta^{2\,\dagger}_\mp \mp \zeta_+ \overline{\zeta^c_+} \wedge \eta^1_+ \overline{\eta^2_\mp} + \zeta_- \overline{\zeta_-} \wedge \eta^1_- \eta^{2\,\dagger}_\pm + \zeta_- \overline{\zeta^c_-} \wedge \eta^1_- \overline{\eta^2_\pm}\ +  \\
	& \ + \zeta_+ \overline{\zeta_-} \wedge \eta^1_+ \eta^{2\,\dagger}_\pm + \zeta_+ \overline{\zeta^c_-} \wedge \eta^1_+ \overline{\eta^2_\mp} \pm \zeta_- \overline{\zeta_+} \wedge \eta^1_- \eta^{2\,\dagger}_\mp \pm \zeta_- \overline{\zeta^c_+} \wedge \eta^1_- \overline{\eta^2_\mp} + \text{c.c.}\ .
\end{split}	
\end{equation}
The presence of the complex conjugates (of all summands) is due to relations such as $\zeta^c_\pm \overline{\zeta_\pm} = - (\zeta_\pm \overline{\zeta^c_\pm})^*$ and $\eta^{1\,c}_\pm \eta^{2\,\dagger}_\pm = -(\eta^{1}_\pm \overline{\eta^2_\pm})^*$.

Since we already know from \eqref{eq:phi} and \eqref{eq:psi} the forms defined by the bispinors along the internal space $M_4$, we just need to compute the bispinors along AdS$_6$, as $\zeta_+ \overline{\zeta_+}$. The structure of these bispinors actually depends on how $\zeta_+$ is chosen. One way to see this is to notice that some of the algebraic relations depend on whether the bilinear $\overline{\zeta_+}\zeta_-$ vanishes or not. A more invariant way to describe the situation is to notice that a pair $\zeta_\pm$ of chiral spinors has the same properties as another pair $\zeta_\pm'$ if they can be related via a Lorentz transformation, $\zeta'_\pm = \Lambda \zeta_\pm$; or in other words if they lie in the same \emph{orbit}. The orbits for SO$(1,5)$ have been studied in \cite[Sec.~2.4.5.2]{bryant-orbits}. Two orbits correspond to the case where either $\zeta_+$ or $\zeta_-$ is zero; these are not compatible with the Killing spinor equation (\ref{eq:KSE}), and are therefore not interesting to us. There is then a one-parameter family of orbits whose stabilizer (i.e.~the little group under the SO$(1,5)$ action) is the abelian group $\rr^4$; each of these orbits has dimension 11. Finally, there is a four-parameter family of orbits whose stabilizer is SU(2); each of these orbits has dimension 12. 

The properties of the forms that one can define from spinor bilinears depend on whether we consider an orbit with stabilizer $\rr^4$ or SU(2). The system in \cite{10d} will give systems of equations which are superficially different for these two types of orbits. However, the original system for supersymmetry is linear in the supersymmetry parameters $\epsilon_a$. So its solution space should be a linear space, which must in fact have dimension 8 (since this is the smallest number of supercharges for a superalgebra in this dimension). Even if two choices of spinor pairs on this linear space might give superficially different systems of equations, eventually these two different systems must agree. So we can choose the spinor pair in such a way as to get the most convenient system of equations. It turns out that this is one of the orbits with $\rr^4$ stabilizer.

To get more concrete, let us decompose the external spinors splitting the external index $\mu$ into a ``lightcone'' part, $a=+,-$, and a four-dimensional Euclidean part, $m=1,\ldots,4$:
\begin{equation}\label{eq:LCgamma}
\gamma^a_{(6)} = \sigma_a \otimes 1_{(4)} = \frac{1}{2}(\gamma^0_{(6)} \pm \gamma^1_{(6)})\ , \quad \gamma^m_{(6)} = \sigma_3 \otimes \gamma^m_{(4)}\ ,
\end{equation}
with $\sigma_\pm = \frac{1}{2}(\pm\sigma_1 + i\sigma_2)$. The matrices $\gamma^\mu_{(6)}$ satisfy the algebra $\cl{1}{5}$ with lightcone metric $\tilde{\eta}^{\mu\nu} = \left[\begin{smallmatrix} 0 & -\frac{1}{2} \\ -\frac{1}{2} & 0 \end{smallmatrix} \right]\oplus \delta^{mn}_{(4)}$, so that $\gamma^{(6)}_\pm = -2 \gamma^\mp_{(6)}$ and $\gamma^{(6)}_m = \gamma^m_{(6)}$. 

Using this decomposition, we choose now a spinor pair of the form
\begin{equation}\label{eq:LCspin}
\zeta_\pm \equiv \begin{pmatrix} 1\\0 \end{pmatrix} \otimes \chi_\pm\ ,
\end{equation}
with $\chi_\pm$ a chiral spinor in four dimensions. This corresponds to an orbit with $\rr^4$ stabilizer. (Orbits with SU(2) stabilizer would correspond to taking $\zeta_+= {1\choose 0}\otimes \chi_+$, $\zeta_-={0\choose 1} \otimes \chi_-$.) One consequence of this (which would not be true for the SU(2) orbit) is that the one-form part of the bilinears $\zeta_+ \overline{\zeta_+}$ and $\zeta_- \overline{\zeta_-}$ coincide; we will call it $z$. It is light-like, and it only has components in the two-dimensional part of the decomposition (\ref{eq:LCgamma}). As for the bilinears in the four dimensions $1,\ldots,4$, they can be evaluated in the same way as those along $M_4$, in terms of two one-forms that we will call $V$ and $W$ and which satisfy exactly the same properties as the forms $v$ and $w$ introduced in \eqref{eq:vw}. 

$z$ and the real and imaginary parts of $V$ and $W$ are independent, and in fact orthogonal. They are not quite a vielbein: if we think of $z$ as of the element of a vielbein in the null direction $-$, we are missing another element in direction $+$. As stressed in \cite{10d}, this cannot be obtained as a bilinear of the supersymmetry parameters; we will see in section \ref{appsub:pair} that the remaining equations in the ten-dimensional system of \cite{10d} require picking such a null vector as an auxiliary piece of data. In conclusion,
\begin{equation}\label{eq:ads6viel}
	 \left\lbrace z=e_-,e_+,\Re V,\Im V,\Re W,\Im W \right\rbrace
\end{equation}
is a vielbein in AdS$_6$. 

We will also define $\Omega_+= - V \wedge W$, $\Omega_- = \bar V \wedge W$, $J_\pm = \pm \frac i2 (V \wedge \bar V \pm W \wedge \bar W)$, just as in (\ref{eq:relform}), (\ref{eq:jpm}) for $M_4$. With all these definitions, we can evaluate
\begin{subequations} \label{eq:extbil}
\begin{align}
& \zeta_\pm \overline{\zeta_\pm} = z \wedge e^{-i J_\pm}\ , \label{eq:zz}\\
& \zeta_+ \overline{\zeta_-} = - z \wedge (V + \ast_4 V)\ , \\
& \zeta_- \overline{\zeta_+} = - z \wedge (\overline{V} - \ast_4 \overline{V})\ , \\
& \zeta_\pm \overline{\zeta^c_\pm} = z \wedge \Omega_\pm\ ,\\
& \zeta_\pm \overline{\zeta^c_\mp} = \mp z \wedge (W \pm \ast_4 W)\ .
\end{align}
\end{subequations}
Specializing to IIB from now on, we can now plug (\ref{eq:extbil}) into \eqref{eq:Phispin}; we have:
\begin{equation}
\begin{split}
	\label{eq:Phiform}
	\Phi_{\text{IIB}} &= e^A \left[ (z \wedge e^{-i J_+}) \wedge \phi^1_+ + (z \wedge e^{-i J_-}) \wedge \phi^1_- \   \right. \\
	&\hspace{1cm}+ z \wedge \Omega_+ \wedge \phi^2_+ + z \wedge \Omega_- \wedge \phi^2_- \    \\
	&\hspace{1cm}  -z \wedge (V+ \ast_4 V) \wedge \psi^1_+ + z \wedge (\overline{V} - \ast_4 \overline{V} ) \wedge \psi^1_- \    \\
	&\left. \hspace{1cm} - \,  z \wedge (W+ \ast_4 W) \wedge \psi^2_+  -  z \wedge (W- \ast_4 W ) \wedge \psi^2_- + \text{c.c.} \right] \ .	
\end{split}	
\end{equation}
This is an odd form, as should be the case for IIB.

To evaluate (\ref{eq:master}), we need to compute the ten-dimensional exterior derivative of $e^{-\phi}\Phi$; schematically, it takes the form:
\begin{equation}\label{eq:Phiextder}
d_H (e^{-\phi}\Phi) = d_H \left(\sum \text{ext} \wedge e^{A-\phi} \, \text{int} \right) =  \sum d_6 \text{ext} \wedge e^{A-\phi}\, \text{int} + (-)^{\deg(\text{ext})} \text{ext} \wedge d_H ( e^{A-\phi} \, \text{int})\ .
\end{equation}
$d_6$ is the differential along the AdS$_6$ coordinates, while $d_H = d_4 - H\wedge$ in the last identity is a combination of the exterior differential $d_4$ along $M_4$ and of the NS three-form $H$ (which only has components along $M_4$). Since we are looking for vacuum solutions to \eqref{eq:master} which are compatible with supersymmetry on AdS$_6$, we need to take the external spinors $\zeta_\pm$ to be the chiral components of a Killing spinor $\zeta$ on this spacetime, i.e.~$\nabla_\mu \zeta = \frac{1}{2} \mu \gamma_\mu \zeta$. The norm of the complex constant $\mu$ (which is proportional to $\sqrt{-\Lambda}$) can be reabsorbed in the warping function $A$; its phase can be reabsorbed by multiplying $\eta^a_\pm$ by $e^{\pm i \theta}$. Hence in what follows we will set $\mu=1$, resulting in the equation (\ref{eq:KSE}) that we already quoted in the main text.  

Exploiting (\ref{eq:KSE}) we can now compute the derivatives of the external forms \eqref{eq:extbil}:
\begin{subequations} \label{eq:derext}
\begin{align}
\label{eq:derext1}
& d_6(\zeta_\pm \overline{\zeta_\pm}) = -2z \wedge (\Re V + 2i \ast_4 \Im V )\ , \\
& d_6(\zeta_\pm \overline{\zeta_\mp}) = \pm 3i z \wedge \Re V \wedge \Im V \pm 5 z \wedge \Re v \wedge \Im V \wedge \Re W \wedge \Im W\ ,\\
& d_6(\zeta_\pm \overline{\zeta^c_\pm}) = -4 z \wedge \ast_4 W\ , \\
& d_6(\zeta_\pm \overline{\zeta^c_\mp}) = \pm 3 z \wedge \Re V \wedge W\ .
\end{align}
\end{subequations}
As an illustration, (\ref{eq:derext1}) is computed as follows:
\begin{align}\label{eq:EXd(ext)}
d_6(\zeta_+ \overline{\zeta_+}) &=\frac{1}{2} \left[ \gamma_{(6)}^{\mu}, \nabla_{\mu} (\zeta_+ \overline{\zeta_+} ) \right]= \frac{1}{4}(\gamma^{\mu}\gamma_{\mu} \zeta_- \overline{\zeta_+}-\gamma^{\mu} \zeta_+ \overline{\zeta_-} \gamma_{\mu}-\gamma_{\mu} \zeta_- \overline{\zeta_+}\gamma^{\mu}+\zeta_+ \overline{\zeta_-} \gamma_{\mu} \gamma^{\mu} ) \nonumber \\
& = \frac{1}{2}(- 3 z \wedge (\overline{V}- \ast_4 \overline{V})- 3 z \wedge (V + \ast_4 V)+ z\wedge (V - \ast_4 V) +z \wedge (\overline{V}+ \ast_4 \overline{V}) )\nonumber \\
&= -2z \wedge (\Re V + 2i \ast_4 \Im V )\ ,
\end{align}
having used the formula 
 $\gamma^{\mu}\omega_k \gamma_{\mu}=(-)^k (D-2k)\omega_k$ for a $k$-form $\omega_k$ in $D$ dimensions.

The left-hand side $d_H (e^{-\phi}\Phi)$ of \eqref{eq:master} then contains only unknown derivatives of the internal forms, since those of the external forms have been traded for the right-hand sides of \eqref{eq:derext}. Once we compute its right-hand side, the complete equation will only involve internal forms and will be valid for any of the sixteen independent components of $\zeta = \zeta_+ + \zeta_-$, as appropriate for an $\mathcal{N}=1$ vacuum in six dimensions.

Before computing the right-hand side of \eqref{eq:master}, namely $-(\tilde{K} \wedge + \iota_K) F$, we will look at the simpler \eqref{eq:symm10d}: as it happens in other dimensions, they imply that the norms of the internal spinors are related to the warping function $A$. Let us see how. First, recall the definitions of $K$ and $\tilde{K}$ \cite{10d}:
\begin{equation}\label{eq:KtildeK}
K=\frac{1}{64}(\overline{\epsilon}_1 \gamma^{(10)}_M\epsilon_1 +\overline{\epsilon}_2\gamma^{(10)}_M\epsilon_2)\, dx^M\ , \quad \tilde{K}=\frac{1}{64}(\overline{\epsilon}_1 \gamma^{(10)}_M\epsilon_1 - \overline{\epsilon}_2 \gamma^{(10)}_M\epsilon_2)\, dx^M\ .
\end{equation}
Plugging in these formulas the decomposition \eqref{eq:10deps}, we obtain:
\begin{equation}\label{eq:KtildeK2}
K = \frac{e^{-A}}{4} z \, (\Vert \eta^1 \Vert^2 + \Vert \eta^2 \Vert^2)\ , \quad \tilde{K} = \frac{e^{-A}}{4} z \, (\Vert \eta^1 \Vert^2 - \Vert \eta^2 \Vert^2)\ .
\end{equation}
The external part of the second equation in \eqref{eq:symm10d} gives $e^{-A} d_6 z \, (\Vert \eta^1 \Vert^2 - \Vert \eta^2 \Vert^2) = 0$ (the right-hand side vanishes since $H$ is purely internal). One can explicitly compute $d_6z$, recalling that $z$ is the one-form part of $\zeta_\pm \overline{\zeta_\pm}$; using \eqref{eq:KSE}, one can show that it is nonvanishing. Thus we get:
\begin{equation}\label{eq:6dsymm1}
\Vert \eta^1 \Vert^2 = \Vert \eta^2 \Vert^2\ .
\end{equation}
Hence $K=\frac{e^{-A}}{2} z\, \Vert \eta^1 \Vert^2$ and $\tilde{K}=0$. On the other hand, the first equation in \eqref{eq:symm10d} says that $K$ is a Killing vector with respect to the ten-dimensional metric $g$: its external part says that $z$ is Killing with respect to $g_{\mathrm{AdS}_6}$ (this is obvious, since $z$ is a bilinear constructed out of Killing spinors), while its internal part implies $\partial_m \left( \frac{e^{-A}}{2} \Vert \eta^1 \Vert^2 \right) = 0$, which upon integration gives
\begin{equation}\label{eq:6dsymm2}
\Vert \eta^1 \Vert^2 = e^A\ ,
\end{equation}
where without loss of generality we have set to one a possible integration constant. 
Putting \eqref{eq:6dsymm1} and \eqref{eq:6dsymm2} together we get \eqref{eq:norms}. Moreover $K=z/2$.
Recalling \eqref{eq:10dflux} we now have:
\begin{align}
-(\tilde{K}\wedge + \iota_K) F_{(10)} &= -\iota_K(e^{6A}\text{vol}_6 \wedge \ast_4 \lambda F) = - \frac{e^{6A}}{2} \ast_6 z \wedge \ast_4 \lambda F \nonumber \\
&= \frac{e^{6A}}{2} (z \wedge \Re V \wedge \Im V \wedge \Re W \wedge \Im W) \wedge \ast_4 \lambda F\ . \label{eq:fluxpost}
\end{align}

Putting everything together, we can now separate the various terms in (\ref{eq:master}) that multiply different wedge products of the one-forms in (\ref{eq:ads6viel}); since those forms are a vielbein in AdS$_6$, they are linearly independent, and each term has to be set to zero separately. In particular, we see from (\ref{eq:fluxpost}) that the RR flux only contributes to one equation. This gives rise to many equations that can then be arranged in SU(2)$_R$ representations by recalling the definitions \eqref{eq:su2} of the SU$(2)$-covariant forms $\Phi_\pm$ and $\Psi_\pm$. This finally results in the system (\ref{eq:64}). 


\subsection{Redundancy of pairing equations} 
\label{appsub:pair}

We will now show that equations (3.1c,d) in \cite{10d},\footnote{The Clifford action from the left (right) of a ten-dimensional gamma matrix on a $k$-form $\omega_k$ is given by \cite{10d}: $$\gamma^{M}_{(10)}\, \omega_k = (dx^M \wedge + g^{MN}\iota_N) \omega_k\ , \qquad \omega_k\, \gamma^{M}_{(10)} = (-)^k (dx^M \wedge - g^{MN}\iota_N) \omega_k\ .$$}
\begin{subequations}\label{eq:10dpair}
	\begin{align}
	&\pai{\ep{+}{1} \cdot \Phi \cdot \ep{+}{2}}{\gamma^{MN}_{(10)} \left[ \pm d_H(e^{-\phi}\Phi \cdot \ep{+}{2}) + \frac 12 e^{\phi}d^\dagger (e^{-2\phi}\ep{+}{2}) \Phi - F_{(10)}\right]} = 0 \ , \label{eq:10dpair1} \\
	&\pai{\ep{+}{1} \cdot \Phi \cdot \ep{+}{2}}{\left[ d_H(e^{-\phi} \ep{+}{1} \cdot \Phi) - \frac 12 e^{\phi}d^\dagger (e^{-2\phi}\ep{+}{2}) \Phi - F_{(10)}\right]\gamma^{MN}_{(10)}} = 0 \ , \label{eq:10dpair2}
	\end{align}	
\end{subequations}
are completely redundant when specialized to $\ads$ solutions in IIB, i.e.~they are automatically satisfied by the expressions for bispinors and fluxes we found in section \ref{sec:gen}. Since the analysis of the case at hand is similar to the ones presented in \cite{10d} and \cite{rosa} (for four- and two-dimensional Minkowski vacua respectively), we will briefly describe the main computations and point out the novelties arising for an AdS vacuum.

Firstly, we need to choose the vectors $\ep{+}{a}$. Intuitively, these auxiliary vectors are needed because the form $\Phi$ is not enough by itself to specify a vielbein; for more details, see \cite{10d}. The $\ep{+}{a}$ can be chosen quite freely, provided they satisfy the constraints
\begin{equation}
\label{eq:epconstraints}
\ep{+}{a}^2 = 0 \ , \quad \ep{+}{a} \cdot K_a = \frac 12\ .
\end{equation}
Since $K_1 = K_2 = K = \frac12 z$ has only external indices, we will set
\begin{equation}\label{eq:e+12}
	\ep{+}{1} = \ep{+}{2} \equiv e_+ \ ,
\end{equation}
and we will consider $e_+$ to be purely external as well. This is just the one-form that in (\ref{eq:ads6viel}) we had to leave undetermined; as we anticipated there, it is an auxiliary piece of data and cannot be determined as a bilinear of $\zeta_\pm$. For Minkowski vacua, $K$ is a constant vector, and one can then simply take $e_+$ to be constant too. In AdS, however, the requirement that $K$ be a Killing vector does not imply that it is constant, and hence there is no reason to have $e_+$ constant either. However, we will argue that $e_+$ can be chosen in such a way to at least make the $d_6^\dagger e_+$ terms in (\ref{eq:10dpair}) vanish. To this end, let us first define the spinors $\tilde{\zeta}_\pm$ along the lines of \eqref{eq:LCspin}:
\begin{equation}\label{eq:auxiliaryspinors}
 \tilde{\zeta}_\pm \equiv \begin{pmatrix} 0\\1 \end{pmatrix} \otimes \chi_\pm\ ,
\end{equation}
and the one-form
\begin{equation}\label{eq:epdefinition}
 e_+ \equiv (\tilde{\zeta}_+ \overline{\tilde{\zeta}_+})_{\text{one-form}} \propto \overline{\tilde{\zeta}_+}\gamma_\mu^{(6)} \tilde{\zeta}_+ \ dx^\mu \ ,
\end{equation}
which satisfies $e_+^2=0$, $e_+ \cdot K \neq 0$; thus, by appropriate rescaling, taking (\ref{eq:e+12}) and (\ref{eq:epdefinition}) will indeed satisfy (\ref{eq:epconstraints}). 
Since (\ref{eq:auxiliaryspinors}) now also satisfies the Killing spinor equations (\ref{eq:KSE}), $d_6^\dagger e_+$ vanishes. 

Another difference with respect to the Minkowski case comes from the term $d_H (e^{-\phi} \Phi \cdot e_+)$. Using the formula $\left\{d, \cdot\, e_+ (-)^{\mathrm{deg}}  \right\} = e^{-A} \partial_+ + dA \wedge e_+ \cdot $, we can write it as 
\begin{equation}
 	d_H (e^{-\phi} \Phi \cdot e_+) = (d_H(e^{-\phi} \Phi)) \cdot e_+ -  e^{-\phi} dA \wedge e_+ \cdot \Phi - e^{-(A+ \phi)} \partial_+  \Phi \ .
\end{equation}
As usual, the first term on the right hand side vanishes inside a pairing,\footnote{This is because $e_+^2=0$. Just replace $C$ with $(d_H (e^{-\phi}\Phi)) \cdot e_+$ in the formula \cite[Sec.~B.4]{10d} $$\pai{e_+ \cdot \Phi \cdot e_+}{C} = - \frac{(-)^{\deg(\Phi)}}{32} \overline{\epsilon_1} e_+ C e_+ \epsilon_2\ .$$} while the last one does not (contrary to the Minkowski case), and we must evaluate it. Since $\partial_+ \Phi = \delta^{+\mu} \nabla_\mu \Phi = \delta^{+\mu} \nabla_\mu (\epsilon_1 \overline{\epsilon_2})$, we can use the decomposition \eqref{eq:10deps} and the equations \eqref{eq:KSE} to conclude that
\begin{equation}
\partial_+ \Phi =\frac 12  e_+ \cdot \hat{\Phi} +  \ldots \ , 
\end{equation}
where the dots denote terms that vanish in the pairing in (\ref{eq:10dpair1}), and where we defined 
\begin{equation}
	 \hat \Phi \equiv (\hat{\epsilon}_1 \overline{\epsilon_2}) \ ,\qquad
	\hat{\epsilon}_1 \equiv \zeta_- \eta^1_+ + \zeta_-^c \eta^{1\,c}_+ + \zeta_+ \eta^1_- + \zeta_+^c \eta^{1\,c}_-\ .
\end{equation}

To sum up, for type IIB $\ads$ vacua we can rewrite \eqref{eq:10dpair1} as
\begin{equation}
\pai{e_+ \cdot \Phi \cdot e_+}{\gamma^{MN}_{(10)} \left[  e^{-\phi} dA \wedge (e_+ \cdot \Phi) + \frac{ e^{- (A + \phi)}}{2} e_+ \cdot \hat{\Phi} - 2 F\right]} = 0 \ ;
\end{equation}
to rewrite the flux term we have made use of the formula 
\begin{equation}
\pai{e_+ \cdot \Phi \cdot e_+}{F_{(10)}} = 2 \pai{e_+ \cdot \Phi \cdot e_+}{F}\ .
\end{equation}

From now on the analysis parallels the one for Minkowski vacua, and we will not repeat it here. Specializing (\ref{eq:10dpair1}), (\ref{eq:10dpair2}) to the case $M=m$, $N=n$ does not give any equations; specializing them to the cases $M=\mu$, $N=\nu$ and $M=m$, $N=\nu$ gives\footnote{As a curiosity, notice that (\ref{eq:pa3}) can also be written as 
\begin{equation}
	\sqrt{g} *\left((\Phi^0_+ - \Phi^0_-)\wedge \lambda(F)\right) = - e^{A-\phi} d A\ . 
\end{equation}
}
\begin{subequations}\label{eq:leftpair}
\begin{align}
& \pai{\Psi^{0}_+ + \Psi^0_- }{F} = e^{-\phi}  \ ,  \label{eq:pa1}\\
& \pai{\Psi^\alpha_+ - \Psi^\alpha_- }{F} = 0\ , \label{eq:pa2}\\
& \pai{dx_m \wedge (\Phi^0_+ - \Phi^0_-)}{F} = -  e^{A-\phi} \partial_m A \ , \label{eq:pa3} \\
& \pai{\iota_m (\Phi^0_+ - \Phi^0_-)}{F} = 0 \ . \label{eq:pa4}
\end{align}
\end{subequations}
It can be shown that these equations transform into identities upon plugging in the expressions for the solutions to the system \eqref{eq:64}. This completes the proof of the redundancy of \eqref{eq:10dpair1} and \eqref{eq:10dpair2} for $\ads$ vacua in type IIB.


\section{AdS$_6$ solutions in eleven-dimensional supergravity} 
\label{app:11d}

We will show here that there are no AdS$_6\times M_5$ solutions in eleven-dimensional supergravity.\footnote{This conclusion was also reached independently by F. Canoura and D. Martelli.}  This case is easy enough that we will deal with it by using the original fermionic form of the supersymmetry equations, without trying to reformulate them in terms of bilinears as we did in the main text for IIB. 

The bosonic fields of eleven-dimensional supergravity consist of  a metric $g_{11}$ and a three-form potential $C$ with 
four-form field strength $G= d C$. The action is
\begin{equation}
S = \frac{1}{(2\pi)^8 \ell_p^9}\int  R *_{11} 1 - \frac{1}{2}G\wedge *_{11}G - \frac{1}{6}C\wedge G \wedge G\ ,
\end{equation}
with $\ell_p$ the eleven-dimensional Planck length. 

We take the eleven-dimensional metric to have the warped product form 
\begin{align}
d s^2_{11} &= e^{2A} d s^2_{\mathrm{AdS}_6} + d s^2_{M_5}\ . 
\end{align}
In order to preserve the SO$(2,5)$ invariance of AdS$_6$ we take the warping factor to be a function of $M_5$, and $G$ to be a four-form on $M_5$. Preserved supersymmetry is equivalent to the existence of a Majorana spinor $\epsilon$ satisfying the equation
\begin{equation}\label{11dSupersymmetryEquation}
\nabla_M \epsilon + \frac{1}{288}\left(\gamma_{M}^{(11)NPQR} - 8 \delta_M^N \gamma_{(11)}^{PQR}\right) G_{NPQR}\, \epsilon = 0\ .
\end{equation}
We may decompose the eleven-dimensional gamma matrices via
\begin{equation}
\gamma^{(6+5)}_\mu = e^A \gamma^{(6)}_\mu \otimes 1\ , \qquad \gamma^{(6+5)}_{m+5} = \gamma^{(6)} \otimes \gamma^{(5)}_m\ .
\end{equation}
Here $\gamma^{(6)}_\mu$, $\mu = 0,\dots,5$ are a basis of six-dimensional gamma matrices ($\gamma^{(6)}$ is the chiral gamma), while $\gamma^{(5)}_m$, $m = 1,\dots,5$ are a basis of five-dimensional gamma matrices. The spinor Anzatz preserving $\mathcal{N} = 1$ supersymmetry in AdS$_6$ is
\begin{equation}\label{11dSpinorAnsatz}
\epsilon = \zeta_+ \eta_+ + \zeta_-\eta_- + {\rm c.c}.
\end{equation}
where $\zeta_\pm$ are the chiral components of a Killing spinor on AdS$_6$ satisfying
\begin{equation}
\nabla_\mu \zeta_\pm = \frac{1}{2} \gamma^{(6)}_\mu \zeta_\mp\ ,
\end{equation}
while $\eta_\pm$ are Dirac spinors on $M_5$.

Substituting \eqref{11dSpinorAnsatz} in \eqref{11dSupersymmetryEquation} leads to the following equations for the spinors $\eta_\pm$:
\begin{subequations}\label{5dSupersymmetryEquations}
\begin{align}	\label{5dSupersymmetryEquation1}
	\frac{1}{2} e^{-A} \eta_\mp \pm \frac{1}{2} \gamma^m_{(5)} \partial_m A \, \eta_\pm + \frac{1}{12} *_5 G_m \gamma^m_{(5)}\eta_\pm &= 0\ ,  \\
	\nabla_m \eta_\pm \pm \frac{1}{4} \ast_5 G_m \eta_\pm \mp \frac{1}{6}*_5 G_n \gamma_m^{(5)} \gamma^n_{(5)}\eta_\pm &= 0 	\label{5dSupersymmetryEquation2}\ .
\end{align}	
\end{subequations}
Using \eqref{5dSupersymmetryEquations} it is possible to derive the following differential conditions on the norms $\eta_{\pm}^{\dagger} \eta_\pm \equiv e^{B_\pm}$ of the internal spinors:
\begin{align}
\ast_5 G &= \mp 6\, d_5 B_\pm \ ,\\
 B_+ &= - B_- + {\rm const}.\ .
\end{align}
We can absorb the constant in a redefinition of $\eta_-$ so that $B_+ = - B_- \equiv B$; thus
\begin{equation}\label{eq:*GB}
\ast_5 G = - 6\, d_5 B\ .
\end{equation}
The equation of motion for $G$ is then automatically satisfied; in absence of sources, the Bianchi identity reads $d_5G = 0$, resulting in $\ast_5 G$ being harmonic. This is in contradiction with $\ast_5 G$ being exact. This still leaves open the possibility of adding M5-branes extended along AdS$_6$, which would modify the Bianchi identity to $d_5G= \delta_{\text{M5}}$. However, we will now show that even that possibility is not realized.

Defining $\tilde\eta_\pm \equiv e^{-B/2} \eta_\pm$ we can rewrite \eqref{5dSupersymmetryEquation2} as
\begin{equation}
\nabla_m \tilde\eta_\pm  \pm \partial_n B\, \gamma_m^n \tilde\eta_\pm = 0\ .
\end{equation}
Upon rescaling the metric $d s^2_{M_5} \rightarrow  e^{-4B} d s^2_{M'_5}$ the equation for $\tilde\eta_+$ becomes
\begin{equation}
\nabla'_m \tilde\eta_+ = 0\ .
\end{equation}
In five dimensions the only compact manifold admitting parallel spinors is the torus $T^5$, so we are forced to set  $d s^2_{M'_5} = ds^2_{T^5}$. Similarly if we rescale the metric $d s^2_{M_5} \rightarrow  e^{4B} d s^2_{M''_5}$ the equation for $\tilde\eta_-$ becomes
\begin{equation}
\nabla''_m \tilde\eta_- = 0\ ,
\end{equation}
so that $d s^2_{M''_5} = ds^2_{T^5}$.\footnote{One might try to avoid this conclusion by setting $\tilde \eta_-$ to zero. However, (\ref{5dSupersymmetryEquation1}) would then also set $\tilde \eta_+$ to zero.} We are thus led to the relation
\begin{equation}
e^{-4B} d s^2_{M'_5} = e^{4B} d s^2_{M''_5}\ .
\end{equation}
Since $d s^2_{M'_5} = d s^2_{M''_5} = ds^2_{T^5}$, this implies $B=0$, and hence $G=0$ (from (\ref{eq:*GB})). This makes the whole system collapse to flat space.


\section{The massive IIA solution} 
\label{app:IIA}

We have shown in appendix \ref{app:11d} that there are no AdS$_6$ solutions in eleven-dimensional supergravity --- and hence in massless IIA. As for massive IIA, it was shown in \cite{passias} that the only solution is the one in \cite{brandhuber-oz}. In this section, we show how that solution fits in the IIA version of the formalism presented in the main text. 

For the bispinors $\Phi$ and $\Psi$, we will keep using the definitions given in section \ref{sec:psp} and the parameterizations given in section \ref{sec:para}. The main difference is the system for supersymmetry, which in IIB was (\ref{eq:64}), and in IIA reads instead
\begin{subequations}\label{eq:64a}
\begin{align}
&d_H \left[e^{3A-\phi} (\Phi_- +\Phi_+)^0 \right] + 2e^{2A-\phi} (\Psi_- - \Psi_+)^0 = 0 \ , \label{eq:aa}\\
&d_H \left[e^{4A-\phi} (\Psi_- +\Psi_+)^\alpha \right] + 3e^{3A-\phi} (\Phi_- - \Phi_+)^\alpha = 0 \ , \label{eq:bcda}\\
&d_H \left[e^{5A-\phi} (\Phi_- +\Phi_+)^\alpha \right] + 4e^{4A-\phi} (\Psi_- - \Psi_+)^\alpha = 0 \ , \label{eq:efga}\\
&d_H \left[e^{6A-\phi} (\Psi_- +\Psi_+)^0 \right] + 5e^{5A-\phi} (\Phi_- - \Phi_+)^0 = - \frac14 e^{6A} \ast_4 \lambda F \ , \label{eq:ha}\\
&d_H \left[e^{5A-\phi} (\Phi_- - \Phi_+)^0 \right] = 0 \ ; \label{eq:ia}\\
& || \eta^1 ||^2 = || \eta^2 ||^2 = e^A \ . \label{eq:normsa}
\end{align}
\end{subequations}

The bispinors $\Phi$ and $\Psi$ can be easily extracted from the supersymmetry parameters: in terms of the vielbein $\{ e^\alpha, e^4 \}$, 
\begin{equation}
	e^\alpha = - w^{-1/6}\frac12 \sin\alpha\, \hat e^\alpha \ ,\qquad e^4=  - w^{-1/6} d \alpha \ , \qquad w \equiv \frac32 F_0\cos\alpha\ ,
\end{equation}
where $\hat e^\alpha$ are the left-invariant one-forms on $S^3$, satisfying 
\begin{equation}
     d \hat e^\alpha = \frac12 \epsilon^\alpha {}_{\beta \gamma} \hat e^\beta \wedge \hat e^\gamma \ ,
\end{equation}
we have
\begin{subequations}
\begin{align}
	\Phi_\pm &= \frac18(\pm 1 - \cos\alpha)\left((1\pm \vol_4){\rm Id}_2+ i\left(\frac12 \epsilon^\alpha{}_{\beta \gamma} e^\beta \wedge e^\gamma \mp e^\alpha \wedge e^4 \right) \sigma_\alpha\right)\ ; \\
	\Psi_\pm &= \frac18 \sin\alpha\,(1 \pm \ast_4)\Big(\mp e^4 {\rm Id}_2 + i e^\alpha \sigma_\alpha\Big)\ ,
\end{align}
being $\sigma_\alpha$ the Pauli matrices.
\end{subequations}

The physical fields then read:
\begin{equation}
	e^\phi =  w^{-5/6} \ ,\qquad e^A= \frac32 w^{-1/6} \ ,\qquad
	ds^2_{M_4}= e^\alpha e^\alpha + e^4 e^4 \ ,\qquad
	F_4 = \frac{10}3 w\, \vol_4\ .
\end{equation}



\bibliography{at}
\bibliographystyle{at}

\end{document}